\documentclass[fleqn,usenatbib]{mnras}

\usepackage{newtxtext,newtxmath}

\usepackage[T1]{fontenc}

\DeclareRobustCommand{\VAN}[3]{#2}
\let\VANthebibliography\thebibliography
\def\thebibliography{\DeclareRobustCommand{\VAN}[3]{##3}\VANthebibliography}

\usepackage{graphicx}	
\usepackage{amsmath}	
\usepackage{bm}         
\usepackage{array}
\usepackage{hyperref}
\usepackage[dvipsnames]{xcolor} 
\usepackage{subfig}

\newcommand\rb{\bm{r}} 
\newcommand\thetab{\bm{\theta}} 

\title[Do mergers increase SF and turbulence?]{Do galaxy mergers increase star formation and turbulence at cosmic noon?}

\author[I. Kanowski et al.]{
I. Kanowski$^{1,2}$\thanks{E-mail: isaac.kanowski@anu.edu.au},
J. T. Mendel$^{1,2}$,
E. Wisnioski$^{1,2}$,
N. M. F\"orster Schreiber$^{3}$,
A. Marchal$^{4,1,2}$,
T. Tsukui$^{1,2}$
\\
$^{1}$Research School of Astronomy and Astrophysics, Australian National University, Canberra, ACT 2611, Australia\\
$^{2}$ARC Centre of Excellence for All Sky Astrophysics in 3 Dimensions (ASTRO 3D), Australia\\
$^{3}$Max-Planck-Institut für extraterrestrische Physik (MPE), Giessenbachstr. 1, D-85748 Garching, Germany \\
$^{4}$Laboratoire de Physique de l’École Normale Supérieure, ENS, Université PSL, CNRS, Sorbonne Université, Université Paris Cité, Observatoire de Paris, \\ F75005, Paris, France
}

\date{Accepted XXX. Received YYY; in original form ZZZ}

\pubyear{\the\year{}}

\begin{document}
\label{firstpage}
\pagerange{\pageref{firstpage}--\pageref{lastpage}}
\maketitle

\begin{abstract}
Mergers and interactions can significantly affect the morphological and dynamical properties of galaxies, however the impact of mergers on turbulence at $z > 1$ has not been observationally constrained. In this work we use the interaction strength parameter $Q_P$ to identify likely interacting and isolated galaxies at cosmic noon ($z \sim 1-2$) within the KMOS\textsuperscript{3D} integral field spectroscopy survey, utilising redshifts from the 3D-HST, CANDELS and UVCANDELS surveys. For $186$ galaxies, we measure deconvolved H$\alpha$ kinematics, including velocity dispersion, using a spatially non-parametric approach to account for observational effects in the dynamically diverse range of galaxies. We compare offsets in H$\alpha$ flux, star formation rate (SFR), dust attenuations, and velocity dispersion of likely interacting galaxies to isolated control galaxies matched in mass and lookback time. We find increased H$\alpha$ fluxes and SFRs in the likely interacting sample at the level of $\sim 0.1$ dex, a similar enhancement to studies of local pairs. In contrast, we find no significant increase in the level of velocity dispersion in interacting galaxies compared to their controls.
The lack of increase in dispersion may reflect a combination of physical and observational factors, including limits to increasing turbulent motions in an already turbulent medium and spectral resolution limits.
\end{abstract}




\section{Introduction}
According to the hierarchical $\Lambda$ Cold Dark Matter ($\Lambda$CDM) paradigm, galaxies grow through two main processes: successive mergers and the accretion of gas.
Mergers are significant events in the life cycles of galaxies $-$ shown to drive both morphological and dynamical diversity across cosmic time \citep[e.g.,][]{Dubois_2016,Sparre_2017,Lagos_2018}.

Interactions of massive ($\log(M_\star/M_\odot) \sim 8-12$) star-forming disc galaxies at $z\lesssim1$ are associated with substantial changes in galaxy star formation rates (SFRs), metallicities and dynamics. Tidal disruptions from close companions are believed to induce inflows of relatively metal-poor gas from the outer regions of galaxies to their more metal-rich centres. These inflows enhance star formation and dilute metallicity throughout the whole disc and most prominently near the galaxy centre \citep{Ellison_2013,Bustamante_2020,Ferreira_2025}.
Studies of thousands of close pairs in the Sloan Digital Sky Suvey (SDSS) ($z \lesssim 0.12$) have found that pairs have up to $\sim0.13-0.25$ dex higher total SFRs than isolated galaxies matched in mass, redshift and environment \citep{Bustamante_2020,Ferreira_2025} and even stronger enhancements when considering only the inner few kiloparsec regions, up to $\sim 0.2-0.3$ dex \citep{Scudder_2012,Ellison_2013}.
Metallicity dillution shows the opposite trend, where close pairs in SDSS have total metallicity offsets of around $-0.035$ dex \citep{Bustamante_2020} and $-0.04$ to $-0.07$ dex offsets within the central few kiloparsecs \citep{Ellison_2013}. Local $z\sim0$ pairs have been found to have flatter metallicity gradients than their matched controls, likely a result of lower metallicity gas flows to the galaxy centres \citep{Kewley_2010}.
Our own galaxy provides a further example, where metallicity dilution has been observed in the age-metallicity relation of Milky Way disc-stars \citep{Ciuca_2024}, believed to be cause by the gas-rich \textit{Gaia}-Sausage-Enceladus major merger.

Both gas and stellar kinematics are also expected to be impacted by interactions. Observations and simulations find that mergers cause complex, non-axisymmetric rotation and velocity dispersion profiles in gas discs \citep[e.g.,][]{Shapiro_2008,Epinat_2009,Bellocchi_2012}. Mergers also increase the intrinsic turbulence of galaxies, likely due to inducing shocks and gas inflows \citep[e.g.,][]{Rich_2015,Fensch_2017,Puech_2019,Perna_2022,Baron_2024}.
\citet{Puech_2019} analysed the velocity dispersion of ionised gas in mergers at $z\sim 0.6$, and, by comparing with simulations to determine merger time evolution, found $\sim1.5-2 \, \times$ increases in velocity dispersion ($\Delta\log\sigma \sim 0.18-0.3$), which correlated with peaks in star formation and gravitational potential perturbations during the merger process.

Galaxy properties are often studied, and observed to change, as a function of pair separation.
SFR enhancements and metallicity dilutions have been observed in pairs with separations as large as $\sim100$ kpc, which may include galaxies which are post-first passage \citep{Scudder_2012,Patton_2013,Bustamante_2020,Ferreira_2025}.
Offsets become stronger as pair separation decreases, particularly at $\lesssim 30$ kpc, and persist or become more substantial post-coalescence \citep{Ellison_2013,Bustamante_2020,Ferreira_2025,ORyan_2025}.
Changes in SFR and metallicity have also been observed in pairs with a wide range of merger mass ratios, from $1/10$ to $1$, with the most significant effects seen in major mergers with mass ratios $> 1/4$ \citep{Ellison_2008,Scudder_2012,Hani_2020}.

Interaction-driven changes in galaxy properties at $z \gtrsim 1$ are less certain, and may change across cosmic time as galaxies evolve.
Galaxies at higher redshift are thicker, more turbulent, gas-rich systems than local galaxies \citep[e.g.,][]{Tacconi_2020,Wisnioski_2025,Yu_2026}.
As such, the primary merger type changes from gas-rich mergers at $z \gtrsim 1$, to gas-poor in the local universe.
In the theoretical framework of \citet{Krumholz_2018}, the primary driver of turbulence also changes as galaxies evolve. Turbulence in local discs is mostly caused by star-formation feedback, whereas turbulence at higher redshift is primarily driven by gas transport within the disc.
SFR density also changes with redshift, peaking at cosmic noon ($z\sim1-2$), and merger rates are higher at earlier times \citep[][and references therein]{Puskas_2025a}.
The impact of mergers on the evolution and scatter in these galaxy properties is also uncertain. Merger-driven SFR and velocity dispersion offsets may be responsible for significant scatter in a range of galaxy scaling relations at $z \gtrsim 1$, including the star-forming main sequence (SFR-M$_\star$), the Tully-Fisher relation (rotational velocity-M$_\star$) and the $\sigma$-SFR relation \citep{Puech_2019}.
Merger-driven turbulence may also contribute to the scatter in observed velocity dispersions at fixed mass and redshift.
Dynamical perturbations from mergers impacting rotational velocity and dispersion profiles may also impact assumptions on the virialised nature of discs, further increasing scatter in relations derived from high redshift samples \citep{Rodrigues_2017}.

Studies of merger-driven SFRs find a range of offsets depending on the sample and method used to measure SFR.
SFR offsets measured using H$\alpha$ have generally been found to be larger than offsets using SFRs measured using Spectral Energy Distribution (SED) fitting codes. A potential explanation for this is that if merger-driven SFR enhancements only occur over short time periods, SED-based SFRs may average over the SFR increases \citep{Horstman_2021}. In contrast, H$\alpha$ emission is sensitive to SFR on shorter timescales, $\sim10$ Myr, and may be able to more accurately capture the merger-driven SFR enhancement. However, SED fitting codes which use non-parametric star-formation-histories may be able to more successfully capture merger-driven SF increases on $\sim 10$ Myr timescales \citep{Calabro_2026}.
Using H$\alpha$-based SFRs of galaxies in the MOSDEF survey at $2 \leq z \leq 2.7$ and $\log(M_\star/M_\odot) \sim 9-11$, \citet{Horstman_2021} find that interacting galaxies in their sample have median SFR offsets of around $0.13-0.24$ dex, depending on the offset calculation method.
Whereas, when using SED-based SFR measurements of the same galaxies, \citet{Horstman_2021} measure lower median SFR offsets of around $0.10-0.16$ dex.
Observations using the \textit{James Webb Space Telescope} (\textit{JWST}) have enabled studies of SFR offsets in pairs beyond cosmic noon out to $z\sim9$.
\citet{Duan_2026} find SED-based SFR offsets in close pairs of $0.25 \pm 0.10$ and $0.26 \pm 0.11$ for galaxies in redshift ranges $4.5 < z < 6.5$ and $6.5 < z < 8.5$ respectively. We note that these enhancements are only observed in the closest pairs (separation $< 20$ kpc) and no significant enhancement was seen at wider separations.
\citet{Calabro_2026} measure SFR offsets of $0.25\pm0.04$ in $\log(M_\star/M_\odot) \sim 7-9$ interacting galaxies at $5<z<8$ identified using morphological indicators\footnote{\citet{Calabro_2026} measure SED-based SFR offsets on a $10$ Myr timescale using non-parametric SFR histories, which may be more comparable to offsets measured from H$\alpha$.}.
In contrast, a number of works have also found no statistically significant offsets in either SFR or specific SFR (sSFR $\equiv$ SFR$/M_\odot$) in galaxy pairs across a range of redshifts \citep[e.g.,][]{Silva_2018,Wilson_2019,Dalmasso_2024}.
Differences likely arise due to galaxy samples, merger selection criteria or SFR measurement method.

A range of simulations have been used to quantify the expected impact of interactions on SFR in galaxies across cosmic time. Idealised hydrodynamical simulations of equal mass ($1$:$1$) major mergers have found that SFR offsets decrease as gas fractions increase \citep{Scudder_2015,Fensch_2017}. \citet{Fensch_2017} found that the SFR enhancements in gas-rich mergers approximating $z\sim2$ massive galaxies ($\Delta \log$ SFR $\lesssim 0.7$) were less than half that of the enhancements in gas-poor $z\sim0$ mergers ($\Delta \log$ SFR $\gtrsim 1$).
\citet{Patton_2020} also found that enhancement of sSFR in close pairs at $z=0.2-1$ was weaker at higher redshifts using the {\tt IllustrisTNG} cosmological simulations {\tt TNG100} and {\tt TNG300} \citep{Marinacci_2018,Springel_2018,Pillepich_2018,Naiman_2018,Nelson_2018}, which included interactions with a range of properties including mass ratios and gas fractions.
Other works using {\tt IllustrisTNG} and {\tt SIMBA} \citep{Dave_2019} cosmological simulations find only minimal changes in SFR increase in post-mergers, out to $z\sim1-2$ \citep{Rodriguez-Montero_2019,Hani_2020}.

Despite the increase in observational studies investigating the effects of mergers on star-formation at $z>1$, there have yet to be comparable studies of galaxy kinematics and turbulence.
In general, simulations find that galaxy mergers at $z \gtrsim 1$ do increase turbulence, though to a lesser extent than at lower redshift \citep[e.g.,][]{Fensch_2017,Mortazavi_2019,Jimenez_2023,Jimenez-Henriquez_2024}.
Merger simulations from \citet{Fensch_2017} found that gas-poor mergers ($\rm{f}_{\text{gas}} = 10$ percent) cause an increase in gas velocity dispersions of $\Delta \log \sigma \sim 0.60$, whereas gas-rich mergers ($\rm{f}_{\text{gas}} = 60$ percent) cause lower increases of $\Delta \log \sigma \sim 0.23$. 
\citet{Jimenez-Henriquez_2024} found a similar trend when comparing turbulence in merging and isolated galaxies in the Evolution and Assembly of GaLaxies and their Environments (EAGLE) simulation \citep{Schaye_2015,Crain_2015}. Mergers at $0.1 < z < 0.5$, had median $\Delta \log \sigma$ offsets which peaked at around $0.16-0.22$ depending on halo mass, whereas mergers at $0.7 < z < 1.3$ had weaker offsets of around $\Delta \log \sigma \sim 0.12$.

In this work we utilise {\tt ROHSA-SNAPD} \citep{Kanowski_2025}, a spatially non-parametric kinematic modelling code to measure the kinematics in a diverse set of star-forming galaxies from the KMOS$^\mathrm{3D}$ survey \citep{Wisnioski_2019}. {\tt ROHSA-SNAPD} is uniquely suited to studying the kinematics of interacting pairs as it can account for the effects of beam smearing without assuming an intrinsic rotation model. The code uses regularisation to enforce smoothly varying kinematics in the underlying model, relying on the assumption that nearby regions of a galaxy have similar intrinsic kinematics. Commonly adopted tools for analysing galaxy kinematics are optimised for regular and axisymmetric rotating disc galaxies \citep[e.g.,][]{Davis_2013,Di_Teodoro_2015,Bouche_2015,Varidel_2019,Price_2021,Lee_2025a} and have not been designed for fitting systems with clearly non-axisymmetric kinematics, particularly merging or interacting galaxies.

In Section \ref{sec:sample_and_methodology} we discuss the compiled galaxy sample analysed in this work and describe the interaction strength parameter we use to identify likely interacting galaxies. In Section \ref{sec:kin_modelling_RSNAPD} we outline our kinematic fitting process and updates to the {\tt ROHSA-SNAPD} code. In Section \ref{sec:results} we present offsets in H$\alpha$ flux, SFR and velocity dispersion of likely interacting galaxies compared to isolated control galaxies matched in mass and lookback time. We discuss potential implications of our results and compare to other merger studies across cosmic time.

In this work we assume a \citet{Chabrier_2003} Initial Mass Function (IMF) and a flat $\Lambda$CDM cosmology with $H_0 = 70$ kms$^{-1}$ Mpc$^{-1}$ and $\Omega_M = 0.3$.

\section{Sample and Methodology}
\label{sec:sample_and_methodology}
To study the kinematic properties of interacting galaxies at cosmic noon, we required a large, mass-complete galaxy catalogue with high-quality redshifts, a subset of which were observed with Integral Field Spectroscopy (IFS). This allowed us to determine the level of interaction between each IFS observed galaxy and the surrounding catalogue galaxies. The 3D-HST Treasury Survey \citep{Brammer_2012,Skelton_2014,Momcheva_2016} was chosen as the parent sample and the KMOS\textsuperscript{3D} survey \citep{Wisnioski_2015,Wisnioski_2019} was chosen as the kinematic IFS subsample.

\subsection{Parent sample}
\label{subsec:parent_sample}
We used source catalogues from the 3D-HST Treasury Survey based on \textit{Hubble Space Telescope} (\textit{HST}) grism observations from five extragalactic fields (AEGIS, COSMOS, GOODS-North, GOODS-South and UDS), primarily of galaxies with redshifts $< 3.5$.
As our goal was to create a high-quality parent catalogue to compare with the KMOS\textsuperscript{3D} sample, we focused on the three southern fields, (COSMOS, GOODS-South and UDS), and applied a number of cuts. The $\rm{{\tt use\_phot}}=1$ quality flag and grism selection criteria of \citet{Kodra_2023} were applied to only include high-quality photometric and grism redshifts across these three fields.
To further increase redshift accuracy, additional spectroscopic and photometric redshift information was compiled from the CANDELS \citep{Kodra_2023}, UVCANDELS \citep{Mehta_2024} and KMOS\textsuperscript{3D} catalogues. Sources were crossmatched with 3D-HST within $0.5$ arcseconds, following \citet{Skelton_2014}.
For sources with multiple spectroscopic redshift measurements, we preferentially selected redshifts measured by the KMOS\textsuperscript{3D} survey.
For sources with multiple photometric redshifts, we first selected photo-zs from UVCANDELS as they include additional photometric bands, then CANDELS {\tt HB4\_z\_weight} redshifts as these are optimised using a combination of methods, and then the original 3D-HST photometric redshifts.
Approximately 20\% of the 3D-HST galaxies across the three fields had updated redshifts that differed by more than $\pm 0.5$ from the original 3D-HST redshifts. We removed these from our catalogue as the stellar masses are likely to have a higher uncertainty given the difference in redshift.

\begin{figure}
    \centering
    \includegraphics[width=\linewidth]{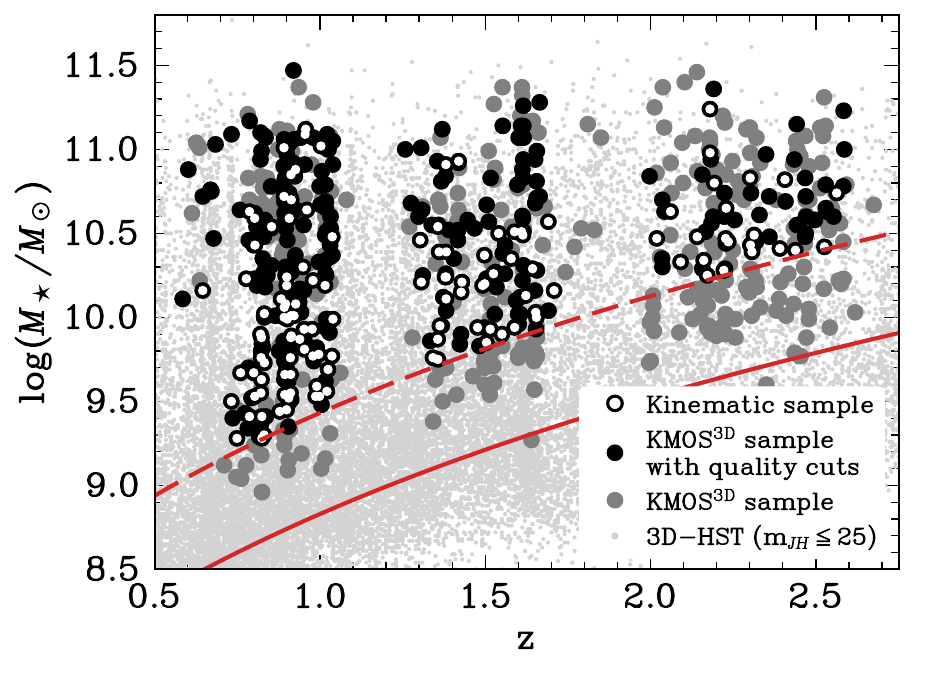}
    \caption{The 3D-HST parent sample, with updated redshift information and quality cuts applied (small grey points), overplotted with the full KMOS\textsuperscript{3D} sample (large grey circles), the KMOS\textsuperscript{3D} sample with quality cuts applied as described in Section \ref{subsec:Kin_sample_introduction} (large black circles), and the final kinematic sample described in Section \ref{subsec:kin_quality_cuts} (large open circles). The solid line describes the fit to the $90$ percent mass completeness limit of the sample. The dashed line is the $4$:$1$ merger mass limit.}
    \label{fig:nintey_percent_completeness_lim}
\end{figure}

We set the magnitude limit of our sample to an observed JH-band magnitude of $m_{\rm{JH}}=25$ because both photometric and grism redshift uncertainties increase towards fainter magnitudes \citep{Bezanson_2016}.
In Figure \ref{fig:nintey_percent_completeness_lim} we show the mass and redshift distribution of the magnitude-limited 3D-HST sample (small grey points). For comparison we also show galaxies from the KMOS\textsuperscript{3D} survey (large open and closed circles) which form our kinematic sample, discussed further in Section \ref{subsec:Kin_sample_introduction}.
We derived the $90$ percent stellar mass completeness limit of the magnitude-limited 3D-HST sample (solid line in Figure \ref{fig:nintey_percent_completeness_lim}) following the procedures of \citet{Marchesini_2009,Pozzetti_2010}. To determine the completeness limit, this approach relies on the assumption that the deeper $26 < m_{\rm{JH}} < 25$ 3D-HST sample includes galaxies with a representative range of mass-to-light ratios.
The $26 < m_{\rm{JH}} < 25$ sample fluxes were scaled to $m_{\rm{JH}} = 25$, with the masses increased accordingly. The scaled masses represented the mass distribution of galaxies at the edge of the $m_{\rm{JH}} = 25$ magnitude limit. The $90$ percent completeness limit of the $m_{\rm{JH}} > 25$ sample was then calculated as a power law fit to the $90^{\rm{th}}$ percentile of the $26 < m_{\rm{JH}} < 25$ scaled mass distribution in $100$ redshift bins between $z = 0.5$ and $2.75$,
\begin{align}
    \log M_\star(z) &= 7.64+3.96\log(1+z) \textnormal{.}
\end{align}
We also calculated a stellar mass limit $4 \times$ greater than the $90$ percent completeness limit (dashed line in Figure \ref{fig:nintey_percent_completeness_lim}). When selecting potentially interacting galaxies, we considered galaxies above this $4 \times 90$ percent completeness limit and looked for potential neighbours above the $90$ percent completeness limit, i.e. our sample is complete for major mergers ($<4$:$1$ merger mass ratio). This is discussed further in Section \ref{subsec:Q_P_parameter}.

We identified five galaxies in the KMOS\textsuperscript{3D} sample which were oversegmented in the 3D-HST catalogue, compared to the contiguous sources observed with KMOS. \textit{JWST} imaging in the redder f277w, f356w and f444w bands was also inspected to confirm the oversegmentation. In these sources, a single galaxy was identified as a number of smaller systems in the 3D-HST catalogue. Oversegmentation generally occurred in edge-on, dusty or clumpy galaxies, where spiral arms or clumps were identified as separate objects. To avoid mis-identifying and oversegmented system as a close pair, we co-added the 3D-HST masses and JH-band magnitudes of the segments to reconstruct a single source.
Two likely spurious sources were also removed from the 3D-HST catalogue as they were located close to a KMOS\textsuperscript{3D} galaxy, but were not visible in any \textit{JWST} or \textit{HST} imaging.

With these cuts applied, the final parent sample consisted of $21 \, 386$ galaxies above the $90$ percent completeness limit, shown in Figure \ref{fig:nintey_percent_completeness_lim}. The sample had spectroscopic, grism and photometric redshift percentages of $30.7$, $5.2$ and $64.1$ respectively.

\subsection{Interaction strength parameter}
\label{subsec:Q_P_parameter}
A common method used to identify galaxy pairs is through maximum projected separation and relative velocity cuts \citep[e.g.,][]{Barton_2000,Patton_1997,Ellison_2008}. However, despite using updated redshifts, the spectroscopic completeness of our parent sample was too low to use this approach. Instead, to identify galaxies which are likely interacting with surrounding galaxies, we used modified versions of the interaction strength parameter \citep{Athanassoula_1984,Dahari_1984,Verley_2007} and redshift Probability Density Function (PDF) pair probability method \citep{Lopez-Sanjuan_2015}.
The interaction strength of a given primary galaxy ($Q_{P}$) is defined as the summation of the weighted interaction strengths of all surrounding galaxies $i$,
\begin{align}
    Q_{P} =& \, \log \sum_i \, w_{iz} \, Q_{iP} \, , \label{Equation:Q_P_full_eqn}
\end{align}

where $Q_{iP}$ is defined as the ratio of tidal to binding force,
\begin{align}
    Q_{iP} \equiv& \frac{F_{tidal}}{F_{bind}} \propto \, \frac{M_i}{M_P} \left( \frac{D_P}{R_{iP}} \right)^3 \, , \label{Equation:Q_iP_derived}
\end{align}
for primary galaxy mass $M_P$ and diameter $D_P$, and mass and projected distance in kiloparsecs to the $i$\textsuperscript{th} neighbouring galaxy $M_i$ and $R_{iP}$.
All galaxy masses were obtained from the 3D-HST catalogue.
We defined $D_P$ as two times the primary galaxy's major-axis effective radius, as measured from \textit{HST} H-band imaging from the \citet{van_der_Wel_2014} catalogue, and converted to kiloparsecs. We note that previous works define $D_P$ using different size measurements, such as the isophotal diameter at $25$ magnitudes arcsecond$^{-2}$ ($\rm{D}_{25}$) or the Petrosian radius which contains $90$ percent of the galaxy light \citep{Karachentseva_1973,Verley_2007,Argudo-Fernandez_2013}. However, we chose to use twice the effective radius because we identified interacting galaxies based on percentiles of $Q_P$ rather than a specific $Q_P$ value, as discussed further below, and therefore only required $D_P$ to reflect the relative size of each galaxy. We also tested calculating $Q_P$ using circularised effective radii and Sersic $\rm{R}_{90}$ sizes and our results did not change significantly.

For neighbouring galaxies with spectroscopic redshifts, we set the redshift weighting $w_{iz}$ to $1.0$ if the redshift difference $z_P - z_i$ implied an absolute relative velocity difference $< 500$ kms$^{-1}$, and $w_{iz} = 0.0$ otherwise. For neighbouring galaxies with grism or photometric redshifts, we calculated $w_{iz}$ as the integral of the redshift PDF within the redshift range which implied a velocity difference of $\pm500$ kms$^{-1}$ from the primary galaxy's spectroscopic redshift. Grism and photometric redshift PDFs were obtained from 3D-HST, CANDELS and UVCANDELS. 
A variety of velocity ranges have been adopted to identify likely interacting pairs in the literature, both limited ranges similar to that adopted here (e.g., $\sim 300-500$ kms$^{-1}$; e.g., \citealt{Patton_2013,Bustamante_2020,Ferreira_2025}), and larger velocity cuts up to $5000$ kms$^{-1}$ or separations in redshift ($\Delta \rm{z} < 0.3$) \citealt{Shah_2022,Duan_2026}).
We discuss the impact of our choice of redshift weighting in Appendix \ref{Appendix:choice_of_redshift_weighting}.
We calculated a $Q_P$ value for every galaxy in our parent and kinematic samples with an available spectroscopic redshift, mass above the $4$:$1$ merger completeness limit and accurate effective radius measurement from \citet{van_der_Wel_2014} (quality flag $> 1$), summing the $Q_{iP}$ relative to every galaxy in the $90$ percent mass-complete parent sample.

We defined the likely interacting and isolated control samples based on percentiles of $Q_P$. Control galaxies had $Q_P < 50^{\rm{th}}$ percentile of $Q_P$ ($Q_P < -4.2$) and likely interacting galaxies had $Q_P > 90^{\rm{th}}$ ($Q_P > -2.4$). We discuss our choice of likely interacting sample further in Appendix \ref{Appendix:Q_P_statistic}.
We chose to define interaction using percentiles of $Q_P$, because we include a redshift weighting term. However, we note that an interaction strength limit of $Q_P \geq -2$, calculated without a redshift weighting, has been used to test for potential tidal impacts from surrounding galaxies \citep{Ubler_2024a}, as this value represents a substantial tidal force at the galaxy edge which is $1$ percent of the binding force \citep{Athanassoula_1984,Verley_2007}.

\begin{figure}
    \centering
    \includegraphics[width=\linewidth]{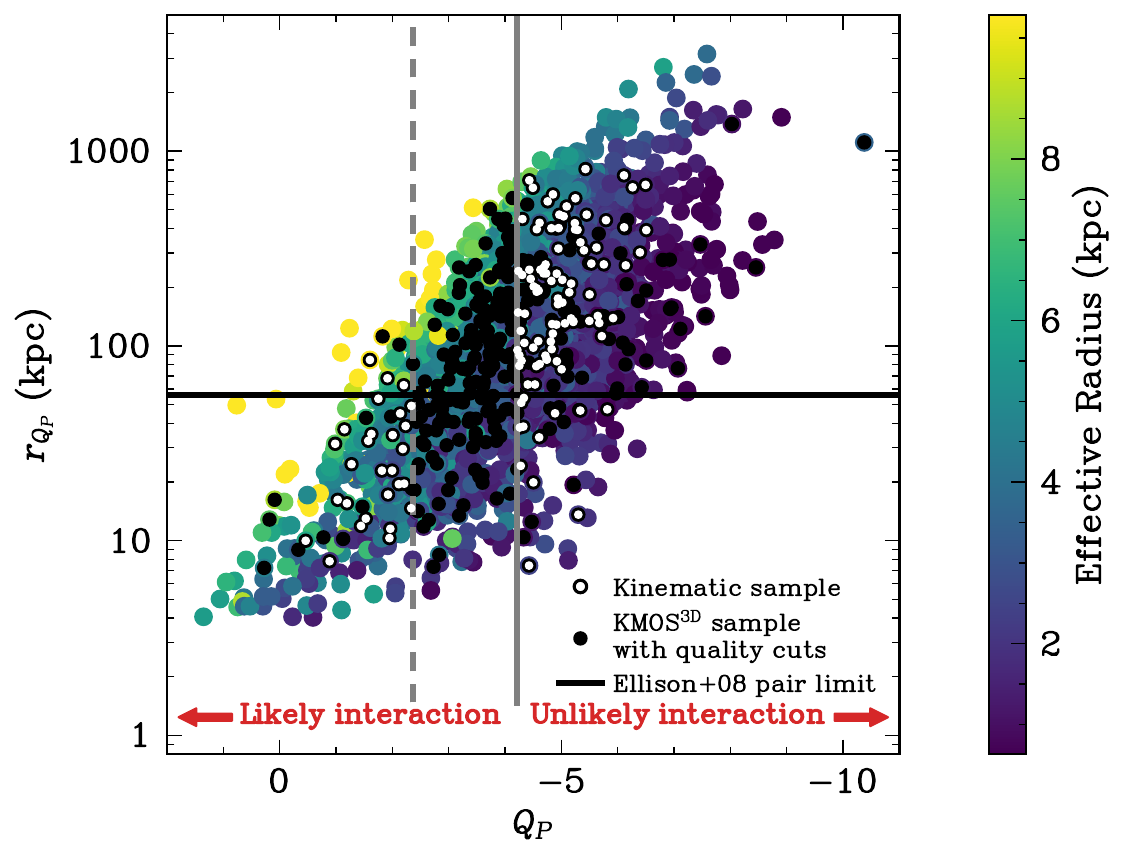}
    \caption{A comparison of $Q_P$ and $r_{Q_P}$ for all galaxies in the spectroscopic redshift sample, coloured by the effective radius of each galaxy. The colourbar spans the $1-99$\textsuperscript{th} percentile of the galaxy effective radii. Closed black circles denote KMOS\textsuperscript{3D} galaxies which pass the quality cuts given in Section \ref{subsec:Kin_sample_introduction} and open circles show the final kinematic sample described in Section \ref{subsec:kin_quality_cuts}. Vertical lines denote the control $Q_P = 50$\textsuperscript{th} percentile limit (solid vertical line) and $Q_P = 90$\textsuperscript{th} percentile likely interacting limit (dashed vertical line). The projected separation pair limit of \citet{Ellison_2008} is included as a horizontal line.}
    \label{fig:Q_P_vs_proj_sep}
\end{figure}

Whilst the low spectroscopic redshift completeness of our sample did not allow pairs to be identified purely with the projected separation method used by other works \citep[e.g.,][]{Barton_2000,Patton_1997,Ellison_2008}, we calculated a similar statistic using the projected separation of each primary galaxy to the galaxy which contributes the largest individual interaction strength, which we denoted as $r_{Q_P}$. The value of 
$r_{Q_P}$ is defined as the $R_{iP}$ to the neighbouring galaxy $i$, which maximises $w_i \, Q_{iP}$. In Figure \ref{fig:Q_P_vs_proj_sep} we compared $r_{Q_P}$ with the interaction strength parameter $Q_P$ for each galaxy in the parent sample with a spectroscopic redshift (coloured circles) to test if likely interacting galaxies defined by $Q_P$ are comparable to pairs identified using the projected separation method. The circles are coloured by the effective radius of each galaxy.
We also overplot KMOS\textsuperscript{3D} galaxies with quality cuts described in Section \ref{subsec:Kin_sample_introduction} as closed black circles and galaxies in the final kinematic sample as open circles, discussed in Section \ref{subsec:kin_quality_cuts}.
We denoted the $Q_P = 50$\textsuperscript{th} percentile control limit (solid vertical line) and the likely interacting $90$\textsuperscript{th} percentile of $Q_P$ (dashed vertical lines). The $80$ h$^{-1}$ kpc projected separation pair limit used by \citet{Ellison_2008} is also shown for comparison (horizontal line). There is a positive correlation between $Q_P$ and $r_{Q_P}$ and scatter in the relation is primarily due to the effective radii of the galaxies, because $Q_P$ is weighted by the size of the primary galaxy, but $r_{Q_P}$ is not.
The majority of likely interacting galaxies have $r_{Q_P}$ separations less than the \citet{Ellison_2008} limit and most control galaxies have $r_{Q_P}$ separations larger than the limit.

\subsection{KMOS\textsuperscript{3D} sample}
\label{subsec:Kin_sample_introduction}
The KMOS\textsuperscript{3D} IFS survey\footnote{KMOS\textsuperscript{3D} data is publicly released at \url{https://www.mpe.mpg.de/ir/KMOS3D}}, used to define our kinematic sample, is a subset of $739$ predominantly star forming 3D-HST galaxies with redshifts $0.6 < z < 2.7$, and stellar masses $9 < \log\left( \textnormal{M}_* / \textnormal{M}_\odot \right) < 11.5$. KMOS\textsuperscript{3D} observations were completed using the K-band Multi Object Spectrograph (KMOS) \citep{Sharples_2013} on the Very Large Telescope (VLT). The data cubes have $0\textnormal{”}.2 \times 0\textnormal{”}.2$ spatial sampling and spectral resolutions between $\textnormal{R}=3 \, 400$ and $4 \, 200$, depending on the band filter used, which corresponds to a spectral resolution, $\sigma_{\textnormal{inst}}$, of around $30$ to $40$ kms$^{-1}$.

We applied a series of quality cuts to the KMOS\textsuperscript{3D} sample, removing $362$ galaxies: $81$ without spectroscopic redshifts, $41$ with updated spectroscopic redshift which differed by more than $0.5$ from the original 3D-HST redshifts, $111$ below the $4$:$1$ mass ratio cut described in Section \ref{subsec:parent_sample}, $108$ likely Active Galactic Nuclei (AGN) using broad-line and NII to H$\alpha$ line ratio classifications from \citet{Forster_Schreiber_2019}, and $20$ with uncertain effective radii measurements (quality flag $> 1$ in \citet{van_der_Wel_2014}\footnote{Galaxies in our sample with quality flags of $1$ in \citet{van_der_Wel_2014} were manually inspected and found to be appropriate.}).
One additional galaxy was removed because its updated spectroscopic redshift meant that H$\alpha$ did not fall within the observed KMOS band.
Applying these quality cuts left $377$ KMOS\textsuperscript{3D} galaxies for which we calculated an interaction strength value, as shown in Figure \ref{fig:Q_P_vs_proj_sep} by open circles. We identified an initial $177$ likely isolated control galaxies ($Q_P < 50^{\rm{th}}$ percentile of $Q_P$) and $40$ likely interacting (($Q_P > 90^{\rm{th}}$ percentile of $Q_P$)) galaxies based on the larger parent sample $Q_P$ percentiles. Further cuts were made to the control and likely interacting samples based on the kinematic fitting results, discussed in Section \ref{subsec:kin_quality_cuts}.

\section{Kinematic modelling with {\tt ROHSA-SNAPD}}
\label{sec:kin_modelling_RSNAPD}
To obtain deconvolved kinematic maps of the KMOS\textsuperscript{3D} galaxies, we used an updated version of the code {\tt ROHSA-SNAPD}\footnote{\url{https://github.com/isaackanowski/ROHSA_SNAPD}}, ``Spatially Non-parametric Approach to PSF Deconvolution using {\tt ROHSA}'' \citep{Kanowski_2025}, which was adapted from the HI emission-line fitting code {\tt ROHSA} \citep{Marchal_2019}. Briefly, {\tt ROHSA-SNAPD} optimises 2D kinematic maps of flux, $\bm{f}(\rb)$, velocity, $\bm{\mu}(\rb)$, and velocity dispersion, $\bm{\sigma}(\rb)$, such that when converted to a 3D model and convolved by both the Point Spread Function (PSF) and data spectral resolution $\sigma_{\textnormal{inst}}$, the convolved model matches well to the observed data.
The option also exists to optimise a single value of velocity dispersion ($\bm{\sigma}$) instead of a full map, which assumes an isotropic velocity dispersion across the whole model, discussed further in Section \ref{subsec:applying_ROHSA_SNAPD}.

Kinematic regularisation is used to ensure that the underlying model is physically meaningful by enforcing that spatial variations in the deconvolved kinematic maps are smooth. The strength of kinematic regularisation is set by hyper-parameters $\lambda_{f}$, $\lambda_{\mu}$ and $\lambda_\sigma$, which weight the relative importance of convolved model goodness-of-fit and kinematic smoothness in the output model. $\lambda_\sigma$ has no effect when optimising a single velocity dispersion value, $\bm{\sigma}$.

\subsection{Updates to {\tt ROHSA-SNAPD}}
Updates have been made to the {\tt ROHSA-SNAPD} codebase to optimise performance with observational data.
We now include the option for a Cauchy likelihood to be adopted as the cost function (default is a Gaussian likelihood), to improve the code's handling of potentially non-Gaussian noise or data impacted by calibration errors (see Appendix \ref{Appendix:likelihoods} for further details).
We also include the ability to mask specific spatial pixels and channels, so that they are not considered in the cost function when comparing the convolved {\tt ROHSA-SNAPD} model to observed data. Masking is used to exclude nearby galaxies and skyline residuals during the kinematic modelling.
The spectral fitting is quite robust to masked channels as each spectrum is modelled by a Gaussian, and velocity regularisation helps to keep fits appropriate, even if parts of emission lines are masked in some channels, as there is often sufficient surrounding data to constrain the fits. However, when spatially masked spaxels are driven purely by regularisation and are completely unconstrained by data. Spatially masked regions in the output model should not be used in further kinematic analysis.

\subsection{Applying {\tt ROHSA-SNAPD}}
\label{subsec:applying_ROHSA_SNAPD}
To prepare the observed data for fitting with {\tt ROHSA-SNAPD}, we subtracted the continuum from each cube following the methodology of \citet{Wisnioski_2019}, before spectrally cropping a $31$ channel wide region centred on the H$\alpha$ emission line (around $67$ {\AA} wide at $z\sim1.5$, corresponding to $1220$ kms$^{-1}$ in velocity space).
We spatially cropped the data to remove pixels with low numbers of exposures, so that more than $50$ percent of the edge pixels in the cropped area were observed with at least $10$ exposures, equivalent to $50$ minutes of on-source observation.
We spectrally masked skyline-dominated spectral channels of each cube, which could otherwise impact the kinematic fitting. We denoted that a given channel was impacted by skylines if more than $75$ percent of the pixels in the channel had RMS errors above the $80$th percentile of the total cube's RMS error.
To ensure the {\tt ROHSA-SNAPD} optimisation algorithm could effectively fit the data, the flux and Root-Mean-Square (RMS) error cubes were scaled so that the integrated per-spaxel fluxes were generally between $1$ and $100$. After fitting the cubes were re-scaled to their original values.

$9$ data cubes contained flux from both the primary galaxy and a nearby pair. $7$ of the galaxies were in the $Q_P > 90$\textsuperscript{th} bin, and $2$ in $Q_P \leq 50$\textsuperscript{th} percentile control bin. We spatially masked the nearby galaxies to reduce their impact on the derived primary galaxy kinematics. Spatial masks were based on segmentation maps from the 3D-HST survey \citep{Skelton_2014}, which were then expanded or contracted based on the observed flux maps to mask the nearby pairs whilst preserving the primary galaxy fluxes as much as possible.

When applying {\tt ROHSA-SNAPD} to the data, we used the fit parameters listed in Table \ref{tab:RSNAPD_params} and all fits were run until they converged. A Cauchy likelihood function was used and bounds on deconvolved flux, velocity and velocity dispersion were set wider than would be expected for galaxies in our sample.

We chose to fit observations with a single deconvolved velocity dispersion value, which assumes that galaxy intrinsic velocity dispersions are isotropic. {\tt ROHSA-SNAPD} fits to simulated KMOS-like data in \citet{Kanowski_2025} suggest that a varying velocity dispersion profile would be poorly constrained for a number of galaxies in our sample, particularly low Signal-to-Noise (S/N) or compact sources.
For this study in particular, fitting a constant deconvolved velocity dispersion, which provides a single dispersion value for each object, also allowed for a clear comparison between the likely interacting and isolated samples, in the same way as parameters such as integrated flux.
Further, deep adaptive optics (AO) supported observations of disc galaxies at cosmic noon do not find significant spatial variations in velocity dispersion profiles \citep[e.g.,][]{Genzel_2011,Forster_Schreiber_2018,Ubler_2019}. However, disrupted sources are more likely to show deviations from an isotropic velocity dispersion profile. We explore potential impacts of the fitting choices in Section \ref{subsec:discussion_vel_disp}.

Within the cost function, the underlying kinematic maps were bilinearly interpolated onto an $n=4$ times finer grid ($0\textnormal{"}.05 \times 0\textnormal{"}.05$ per pixel), before being converted into a convolved model, spatially binned and compared to the observed data. This was done to reduce the impact of sub-pixel velocity gradients on the kinematic deconvolution. As such, we required oversampled resolution PSF models for each galaxy, which we obtained from 2D Moffat models of the observed PSFs, as listed in the KMOS\textsuperscript{3D} fits headers. PSF models were projected onto $4\textnormal{"}.25 \times 4\textnormal{"}.25$ ($85 \times 85$ pixel) grids.

\begin{table}
    \centering
    \begin{tabular}{c|c}
        Parameter & Value \\
        \hline
        Likelihood & Cauchy \\
        constant dispersion & True \\
        oversample factor $n$ & 4 \\
        $\bm{f}(\rb)$ bounds [scaled flux] & ($10^{-6}$,$10^{3}$) \\
        $\bm{\mu}(\rb)$ bounds [km s$^{-1}$] & $\pm400$ km s$^{-1}$ from spectral centre \\
        $\bm{\sigma}$ bounds [km s$^{-1}$] & ($10^{-6}$, $300$) \\
        tol (fit tolerance) & $10^{-10}$ \\
        $\lambda_f$ & $0.0005$ \\
        $\lambda_\mu$ & $2, 4, 8$ \\
    \end{tabular}
    \caption{The parameters used to apply {\tt ROHSA-SNAPD} to the KMOS\textsuperscript{3D} galaxies.}
    \label{tab:RSNAPD_params}
\end{table}

\subsection{Choice of regularisation hyper-parameters}
\label{subsec:choice_regularisation}

When applying {\tt ROHSA-SNAPD} to the data, we set the flux regularisation at $\lambda_f = 0.0005$ and tested three velocity regularisation values $\lambda_\mu = 2$, $4$ and $8$.
The choice of regularisation hyper-parameters $\lambda_f$ and $\lambda_\mu$ can have significant impacts on the deconvolved model. Flux regularisation reduces the impact of noise during fitting and enforces spatial coherence in the model. Velocity regularisation enforces the level of smoothness in the underlying velocity field to both reduce the impact of noise and alleviate degeneracies between the velocity and velocity dispersion maps.
$\lambda_f$ and $\lambda_{\mu}$ values can be approximated for each galaxy in the sample using, for example, the increase in $\chi^2$ \citep{Press_2007} or the L-curve method \citep{Hansen_1992}. However, choosing individual regularisation values is a difficult, computationally expensive task, likely requiring fitting each observation with a range of regularisation values.
Therefore, we instead chose to use the same regularisation values across the entire sample. As we studied galaxy kinematics using a statistical approach, individual kinematic fitting errors due to imperfect regularisation choices were likely averaged out.
Because velocity regularisation often has a larger influence on the deconvolved model than flux regularisation, we chose a single $\lambda_f$ value which was appropriate for the full sample, and applied a small number of reasonable $\lambda_{\mu}$ values to test the sensitivity of our results to velocity regularisation.

\subsection{Monte Carlo resampling}
We used Monte Carlo resampling to derive uncertainties on the {\tt ROHSA-SNAPD} fits. As the RMS error of each pixel was known, we generated a noise cube by sampling a Gaussian distribution for each pixel, with a median of $0$ and standard deviation equal to the pixel RMS.
To avoid decreasing the S/N of the observed data during the resampling, we chose not to add the generated noise cube to the observed data itself. Instead, we assumed that the convolved model of the {\tt ROHSA-SNAPD} fit provided a good estimate of the observed data and created Monte Carlo resamples by adding the generated noise cubes to the convolved model. The assumption of a good convolved model fit is reasonable, as the reduced $\chi^2_R$ of our full sample is $\chi^2_R = 1.33 ^{+ 0.08} _{- 0.07}$ when fitting with $\lambda_f = 0.0005$, $\lambda_\mu = 2$.
Using this method we generated $100$ noise resamples for each data cube and fit them with {\tt ROHSA-SNAPD} to obtain errors on the deconvolved kinematics. Regardless of the velocity regularisation value used to fit the $100$ noise resamples, we used the $\lambda_f = 0.0005$, $\lambda_\mu = 2$ initial fits as they generally provided the best fit to the convolved model, because lower regularisation values give more weight to model goodness-of-fit.

\subsection{Kinemtic quality cuts}
\label{subsec:kin_quality_cuts}

We used the median {\tt ROHSA-SNAPD} model of each galaxy in the KMOS\textsuperscript{3D} sample described in Section \ref{subsec:Kin_sample_introduction}, obtained from fitting $100$ noise resamples of each cube, to compare the kinematics of the likely interacting and control galaxy samples. Deconvolved kinematic maps of flux, line-of-sight velocity and velocity dispersion were returned by {\tt ROHSA-SNAPD} directly. Convolved kinematic maps were calculated by converting the $100$ deconvolved kinematic maps into convolved models and fitting Gaussian functions to each line of sight.
A deconvolved S/N map was calculated for each galaxy by spectrally collapsing the median deconvolved model and observed RMS noise cubes and then dividing the two maps. A convolved S/N map was calculated in the same way, using the convolved model cube.
A single deconvolved flux value was also calculated for each galaxy, as the sum of all spaxels in the median deconvolved flux map with deconvolved S/N $> 1$.

The kinematic and S/N maps were used to make a number of additional cuts to ensure a high quality sample. 
We removed $47$, $46$ and $48$ galaxies from the $\lambda_\mu = 2$, $4$ and $8$ fits respectively which had less than nine spaxels with a convolved S/N $> 1$.
A further $37$, $38$ and $36$ galaxies from the $\lambda_\mu = 2$, $4$ and $8$ fits were removed through visual inspection of the kinematic maps and model residuals, as they were fit poorly by {\tt ROHSA-SNAPD}, were significantly impacted by skyline residuals or had low S/N despite passing the S/N cut mentioned above.
$6$ galaxies were also removed because they had neighbours at very small projected separations. At the KMOS $0\textnormal{"}.2 \times 0\textnormal{"}.2$ spatial sampling it was not possible to separate the neighbouring sources from the primary, even in the deconvolved flux maps, making it impossible to measure kinematics of only the primary galaxy.
$12$, $13$ and $12$ galaxies that were poorly resolved in the $\lambda_\mu = 2$, $4$ and $8$ fits respectively were kept in the sample when fluxes and SFRs of the interacting and control samples, but were removed when comparing velocity dispersions. Poorly resolved galaxies were visually identified using their convolved kinematic maps and PSF Full Width Half Max (FWHM) sizes, and were generally smaller than twice the PSF FWHM.

\begin{figure*}
    \centering
    \begin{tabular}{m{0.26\linewidth} m{0.68\linewidth}}
    \centering\includegraphics[width=\linewidth]{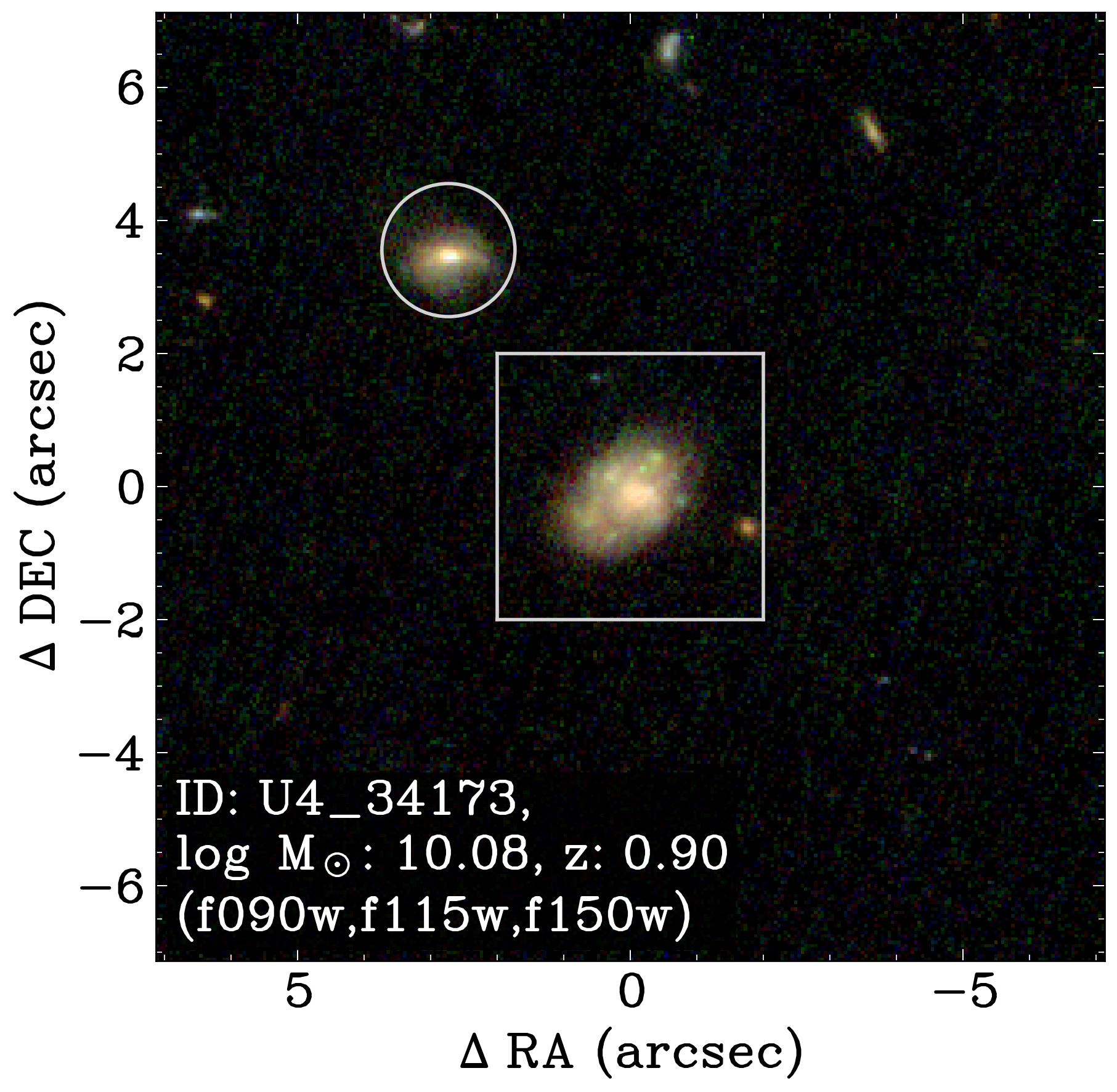} &
    \centering\includegraphics[width=\linewidth,trim={25cm 0cm 0cm 0cm},clip]{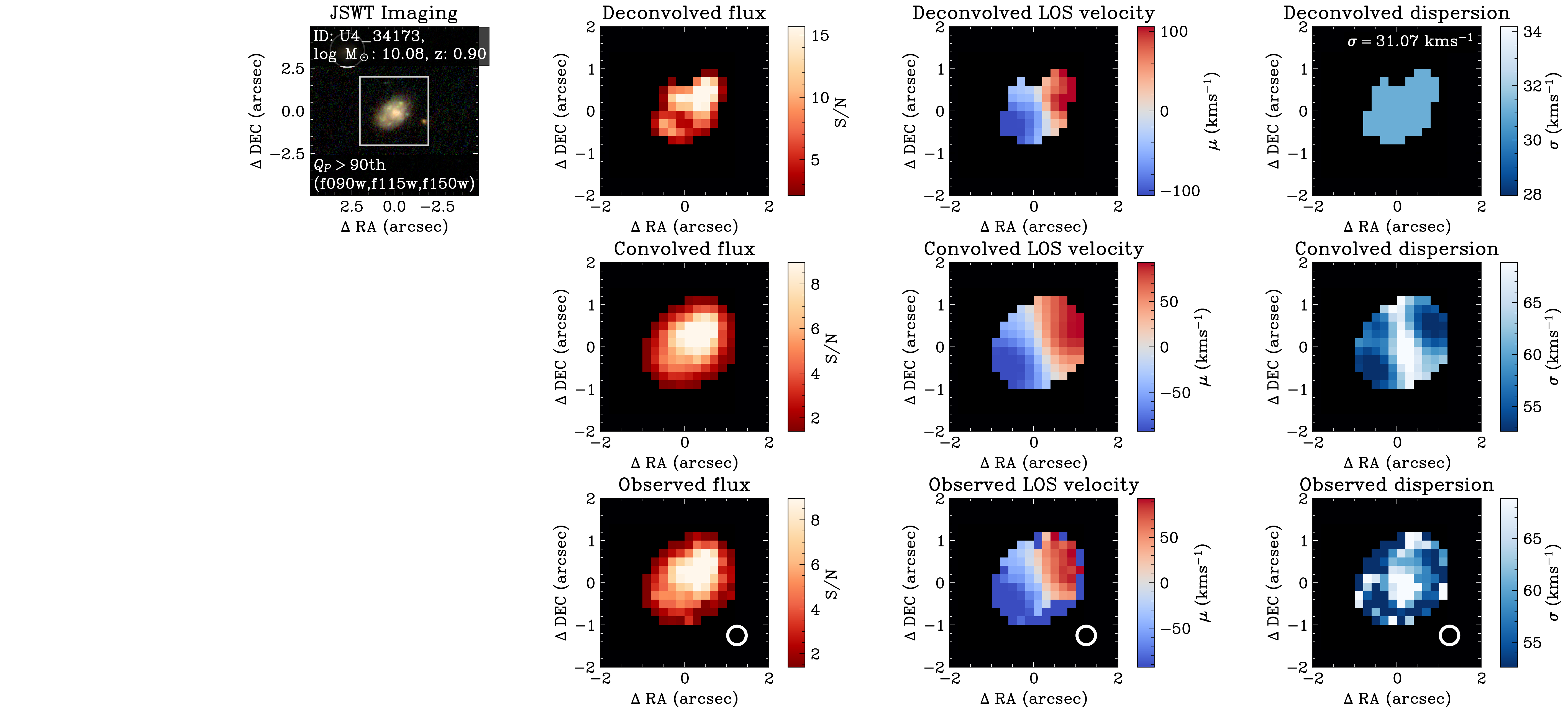}
\end{tabular}
    \caption{This figure shows the galaxy {\tt U4\_34173}. The left-most panel shows the galaxy in a 3-colour \textit{JWST} image (f090w, f115w, f150w), with the KMOS field of view overplotted as a box. The remaining panels show the results of the {\tt ROHSA-SNAPD} modelling, where each map is the median fit from $100$ noise resamples. As the KMOS\textsuperscript{3D} cubes were cropped to remove noisy edge pixels, as discussed in Section \ref{subsec:applying_ROHSA_SNAPD}, we pad the edges of the maps to the KMOS field of view for clear comparison with the \textit{JWST} imaging. The top row shows deconvolved flux, LOS velocity and velocity dispersion. The bottom row shows convolved flux, LOS velocity and dispersion.
    Deconvolved and convolved kinematic maps shows pixels with S/N $> 1$, calculated using deconvolved and convolved flux maps respectively, as discussed in Section \ref{sec:results}.
    The FWHM ellipse of the observed PSF is included on the kinematic maps in the bottom row.
    The colourbar bounds for the deconvolved and convolved {\tt ROHSA-SNAPD} kinematic maps are calculated separately using non-masked spaxels, and the observed maps use the same bounds as the convolved.
    The flux map colourbar spans the $10$ to $90$\textsuperscript{th} percentiles. The line-of-sight velocity map is set to $\pm$ the $10$ to $90$\textsuperscript{th} percentile, whichever has the largest magnitude. All line-of-sight velocity maps are centred around $v_{\rm{max}}+v_{\rm{min}}/2$ of the deconvolved map.
    The deconvolved velocity dispersion is a single value for the full map, so the colourbar is centred on this value, and is also listed in the top right corner of the panel. The convolved and observed velocity dispersion colourbars span the $10$ to $90$\textsuperscript{th} percentiles of the convolved map.}
    \label{fig:merger_kin_maps_example}
\end{figure*}

In Figure \ref{fig:merger_kin_maps_example} we show an example fit to a likely interacting galaxy {\tt U4\_34173}. The far left panel shows a three-colour (f090w, f115w, f150w) \textit{JWST} image of the galaxy obtained from the DAWN \textit{JWST} Archive\footnote{DAWN \textit{JWST} archive: \url{https://dawn-cph.github.io/dja/}}, with an overplotted white box showing the $4\textnormal{"} \times 4\textnormal{"}$ field of view achieved by the observation pattern of the KMOS\textsuperscript{3D} survey. The white circle indicates the neighbouring galaxy that contributed the largest individual interaction strength $Q_{iP}$, with a mass ratio of $1$:$4.7$, at a projected separation of $34.8$ kpc. The remaining panels show the deconvolved and convolved {\tt ROHSA-SNAPD} maps and observed kinematics. The observed galaxy kinematics were obtained by fitting Gaussian functions to each spaxel of the observed data cube, accounting for the RMS error. Deconvolved maps were masked to only include spaxels with deconvolved S/N $> 1$, while the convolved and observed maps show spaxels with convolved S/N $> 1$.
The deconvolved flux map shows multiple regions of star formation, which is similar to the structure seen in the \textit{JWST} imaging. The deconvolved velocity and velocity dispersion are indicative of a disc-like system. 
We show the results of all other likely interacting galaxies in Appendix \ref{Appendix:all_Q_P_g_90_fits}. Most systems were reasonably disc-like, though a number of galaxies showed disrupted morphologies or velocity profiles.

\begin{table}
    \centering
    \begin{tabular}{c|c|c|c}
        $Q_P$ bin & $\lambda_\mu = 2$ & $\lambda_\mu = 4$ & $\lambda_\mu = 8$ \\
        \hline
        $Q_P < 50$\textsuperscript{th} & $99$ $(92)$ & $100$ $(92)$ & $99$ $(92)$ \\
        $Q_P > 90$\textsuperscript{th} & $28$ $(23)$ & $27$ $(22)$ & $28$ $(23)$ \\
        \hline
        Total & $127$ $(115)$ & $127$ $(114)$ & $127$ $(115)$
    \end{tabular}
    \caption{The final size of the kinematic control and likely interacting samples, for fits with different $\lambda_\mu$ values. Values in brackets in each cell show the number of galaxies which meet the well-resolved criteria discussed in Section \ref{subsec:kin_quality_cuts}.}
    \label{tab:kin_sample_sizes}
\end{table}

When fit with $\lambda_f = 0.0005$ and $\lambda_\mu = 2$, $58$ of the $120$ ($48.3$ percent) well-resolved galaxies returned a median deconvolved velocity dispersion below $10$ kms$^{-1}$, an approximate intrinsic velocity dispersion lower limit which would be expected due to thermal motions \citep{Shields_1990,Ubler_2019}. A majority of the $69$ galaxies had dispersions at the {\tt ROHSA-SNAPD} velocity dispersion lower bound of $10^{-6}$ channels, $\sim0$ kms$^{-1}$. These low fits implied that, given the instrument resolution and S/N of the data, the observed dispersion was almost entirely reproduced through the combination of the data spectral resolution $\sigma_{\textnormal{inst}}$ and smearing of the velocity map. There were similar fractions of $<10$ kms$^{-1}$ dispersion fits for the $\lambda_\mu = 4$ and $\lambda_\mu = 8$ fits, $42.0$ and $26.7$ percent respectively. Because increased velocity regularisation enforces smoother velocity maps, the fit velocity dispersions increase, reducing the number of fits below $10$ kms$^{-1}$ for $\lambda_\mu = 4$ and $\lambda_\mu = 8$ compared to $\lambda_\mu = 2$.
To include galaxies with $<10$ kms$^{-1}$ velocity dispersion fits in our sample, we adopt an upper limit dispersion equal to the data spectral resolution $\sigma_{\textnormal{inst}}$ ($\sigma_{\rm{inst}} \sim 30-40$ kms$^{-1}$), which is an approximate point above which we would expect to reliably recover galaxy velocity dispersions using {\tt ROHSA-SNAPD}.

The final sample for each regularisation value is given in Table \ref{tab:kin_sample_sizes}. We refer to these galaxies as our kinematic sample. Below we describe the final quality cuts applied.

\section{Results}
\label{sec:results}
With our final sample we compare global properties of $130$ galaxies ($120$ well resolved) across the redshift range $0.6<z<2.6$. 
To compare a given parameter (e.g., H$\alpha$ flux) between the control and likely interacting samples, we calculated offsets between each likely interacting galaxy and a sample of control galaxies with similar physical properties, using a similar method to other merger studies \citep[e.g.,][]{Patton_2011,Scudder_2012}.

For each galaxy in the high $Q_P$ "likely interacting" sample, we identified galaxies in the control sample with lookback times and stellar masses within $\pm0.3$ Gyr and $\pm0.1$ dex of the likely interacting galaxy.
We chose these mass and lookback time bounds as a balance between minimising the shift in the normalisation of the star-forming-main-sequence ($\lesssim 0.1$ dex) and ensuring there were sufficient control galaxies to match to.
We then randomly chose one of the matched controls to calculate the offset in a parameter (e.g., $\Delta \log$ H$\alpha$ flux) between the likely interacting galaxy and the control.
If no matches existed, we expanded the mass and redshift bounds by $\pm0.1$ Gyr in lookback time and $\pm0.1$ dex in stellar mass until at least one match occurred. Once calculated for all likely interacting galaxies, we saved the median offset (e.g., median $\Delta \log$ H$\alpha$ flux). We repeated this process $200$ times, each iteration randomly choosing another control galaxy for each $Q_P > 90^{\rm{th}}$ galaxy, to obtain a distribution of median offsets of the parameter. We then jackknife resampled the distribution, as we found during testing that the removal of single galaxies can sometimes alter the measured offsets, though the impact of jackknifing was generally minimal.
This process was repeated to match the control sample to itself.

\begin{figure}
    \centering
    \includegraphics[width=\linewidth,trim={0cm 0cm 0cm 3.3cm},clip]{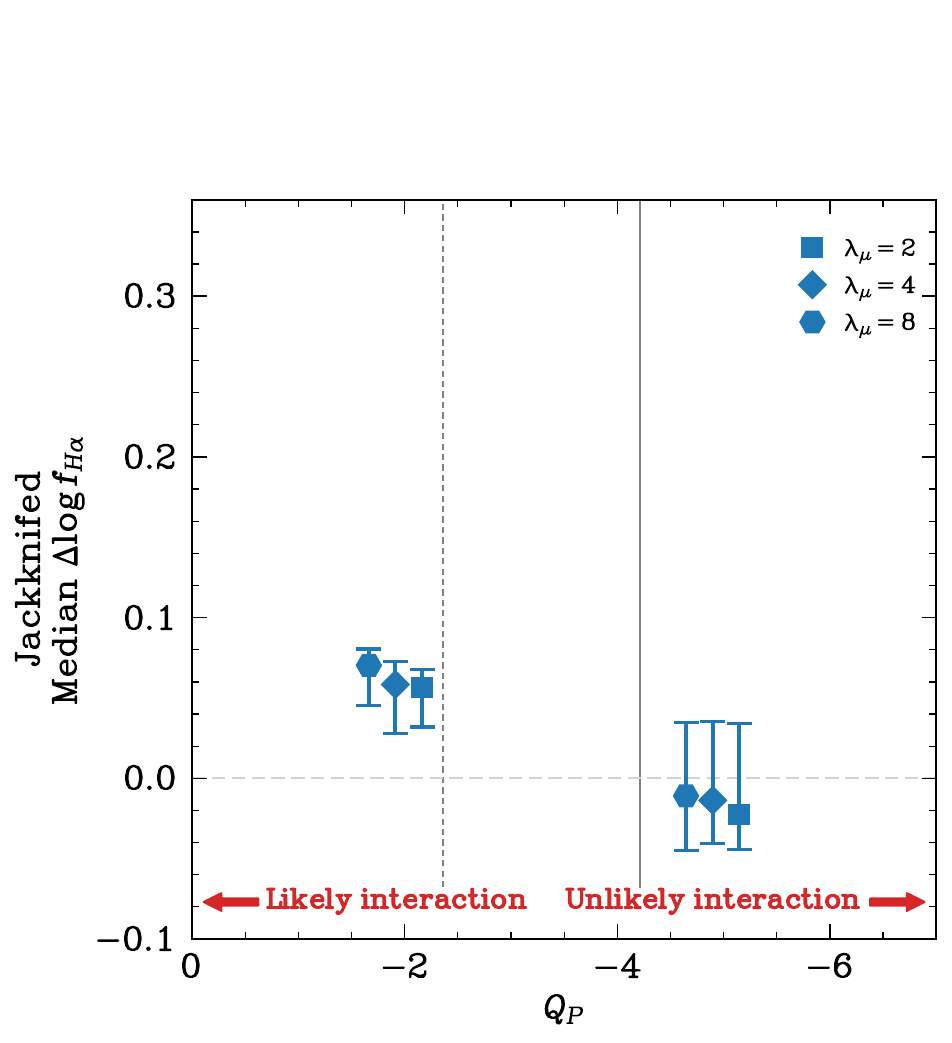}
    \caption{A comparison of interaction strength $Q_P$ with the jackknifed $\Delta \log$ offset in galaxy H$\alpha$ flux. Symbols denote the velocity regularisation value, $\lambda_\mu$, used in the fit. Vertical solid and dashed lines show the control and likely interacting $Q_P$ bounds respectively. Small offsets in $Q_P$ are applied to the different $\lambda_\mu$ results for visual clarity.}
    \label{fig:RSNAPD_results_FHa}
\end{figure}

Figure \ref{fig:RSNAPD_results_FHa} shows the $\Delta \log$ offset in deconvolved H$\alpha$ fluxes for each velocity regularisation $\lambda_\mu$ value. We find that likely interacting galaxies have higher H$\alpha$ fluxes by $0.06^{+0.01}_{-0.03}$ dex ($\lambda_\mu = 4$) than isolated galaxies matched in mass and lookback time. The offsets of different $\lambda_\mu$ value fits are within error of each other, which was expected as velocity regularisation primarily impacts the velocity and velocity dispersion more than flux.

\begin{figure}
    \centering
    \includegraphics[width=\linewidth,trim={0cm 0cm 0cm 3.3cm},clip]{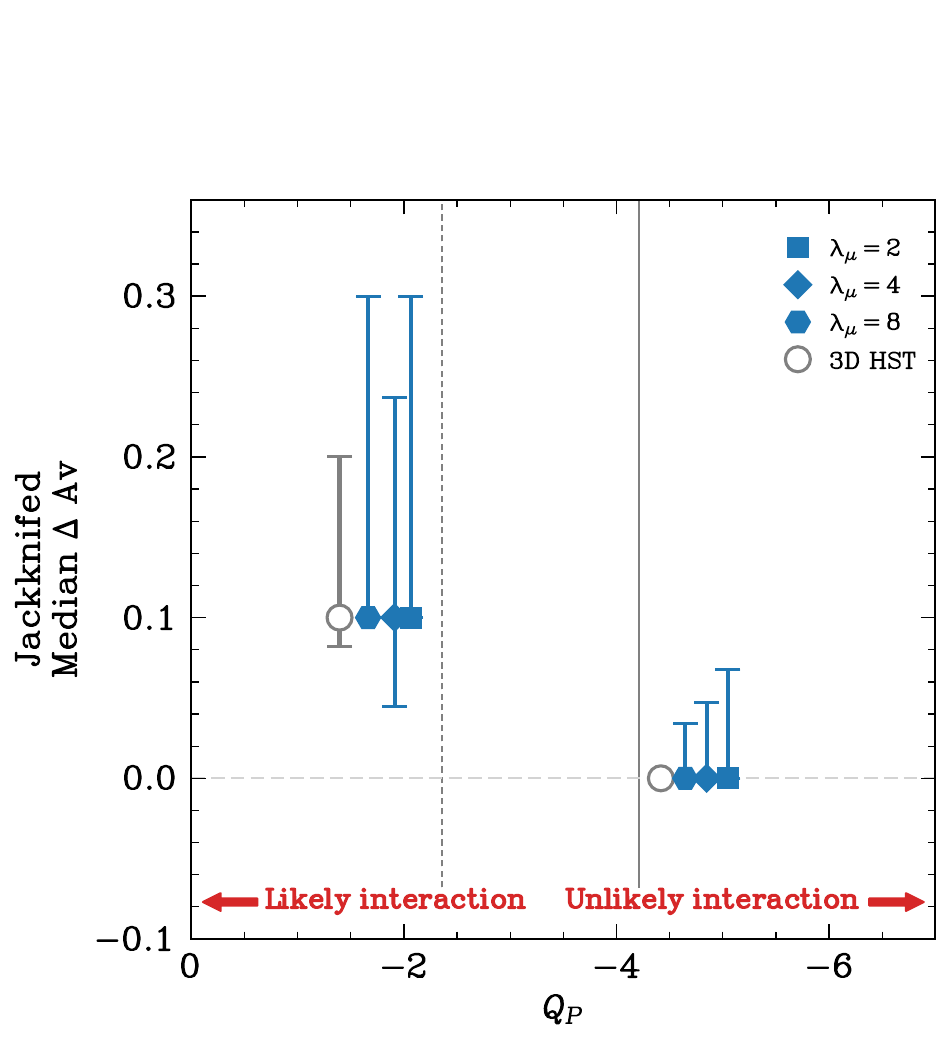}
    \caption{The $\Delta \rm{A}_v$ offset from the control for both the kinematic sample (closed points) and the high-quality star-forming 3D-HST spectroscopic sample (open points). The layout of the figure is similar to Figure \ref{fig:RSNAPD_results_FHa}.}
    \label{fig:RSNAPD_results_Av}
\end{figure}

We converted the deconvolved H$\alpha$ fluxes to star formation rates using the \citet{Kennicutt_1998} relation, shifted to a \citet{Chabrier_2003} IMF using a $1.7 \times$ correction factor and adjusted for attenuation using SED-derived $\rm{A}_v$ factors from \citet{Momcheva_2016} following \citet{Wuyts_2013,Wisnioski_2019}.
Figure \ref{fig:RSNAPD_results_Av} shows that the $\rm{A}_v$ values of the likely interacting galaxies were increased compared to their controls, by $0.10^{+0.14}_{-0.06}$ for the $\lambda_\mu=4$ sample.

We verified the positive attenuation offset in likely interacting galaxies using a high-quality subset of the 3D-HST sample with similar properties to the KMOS\textsuperscript{3D} sample, and found good agreement, a $\Delta A_{\rm{V}}$ offset of $0.10^{+0.10}_{-0.02}$. The high-quality 3D-HST subset included all galaxies which had spectroscopic redshifts in the range $0.6 < z < 2.7$, were above the $4$:$1$ mass ratio limit and were classified as star forming from the UVJ colour-colour criteria of \citet{Whitaker_2011}. We also removed any galaxies which had companions at very close projected separations ($< 2$ arcseconds) to reduce the impact of unidentified oversegmented sources.

\begin{figure*}
    \centering
    \includegraphics[width=0.49\linewidth,trim={0cm 0cm 0cm 3.3cm},clip]{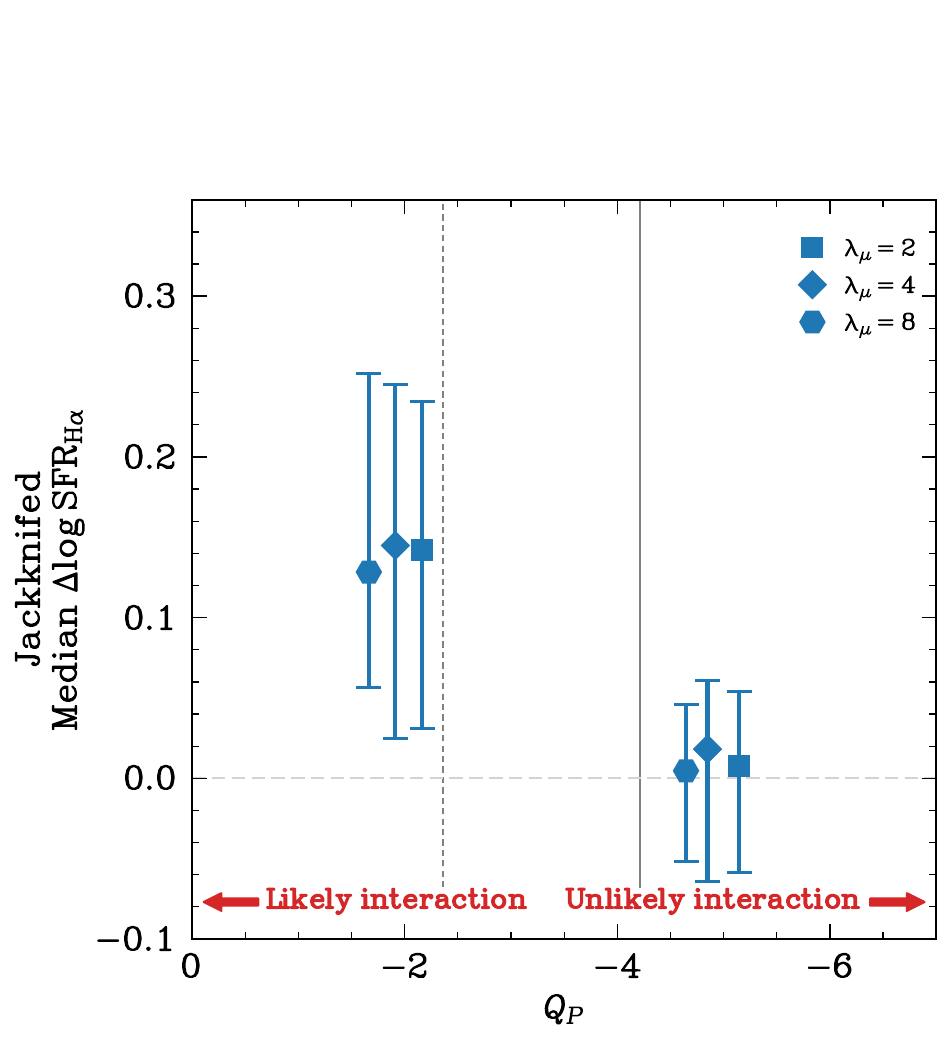}
    \includegraphics[width=0.49\linewidth,trim={0cm 0cm 0cm 3.3cm},clip]{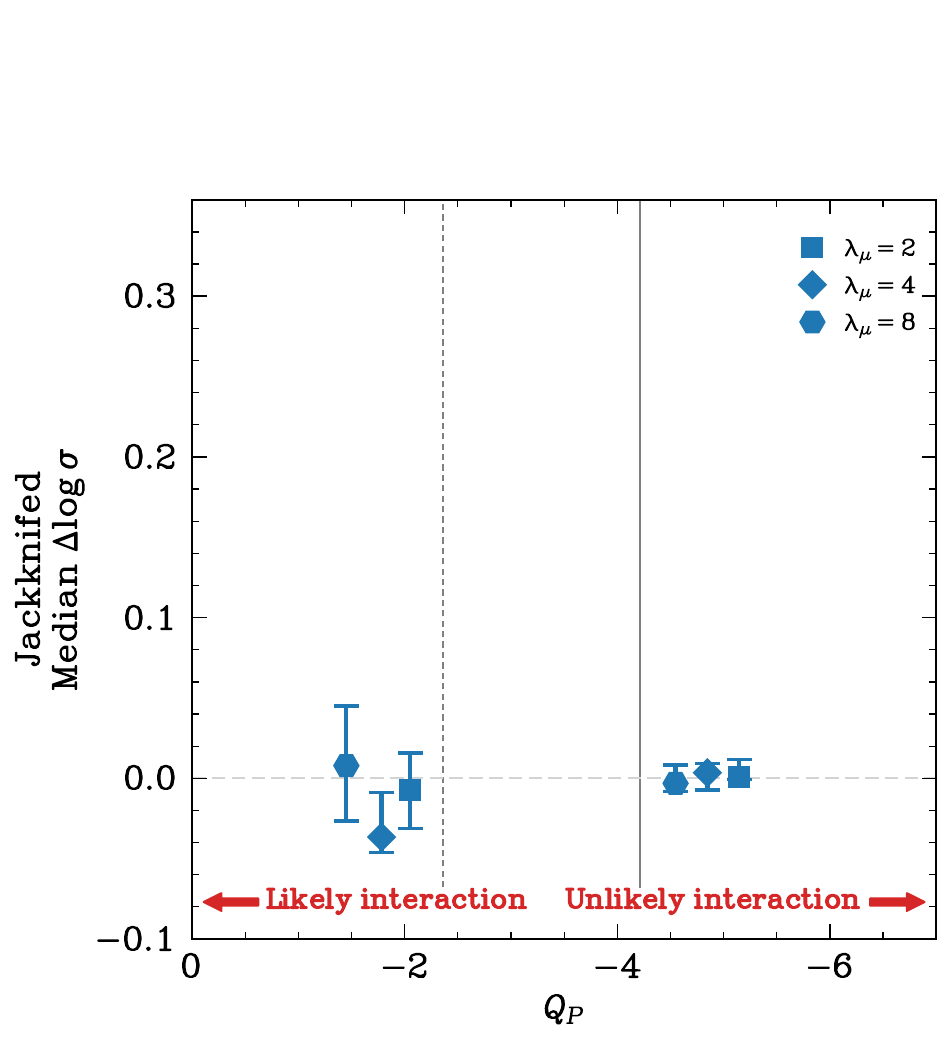}
    \caption{The two panels compare interaction strength with the $\Delta \log$ offset from the control of galaxy H$\alpha$ SFR (left panel) and deconvolved velocity dispersion (right panel). The velocity dispersion offsets are calculated using only well-resolved galaxies, as described in Section \ref{subsec:kin_quality_cuts}. The layout of each panel is similar to Figure \ref{fig:RSNAPD_results_FHa}.}
    \label{fig:RSNAPD_results}
\end{figure*}

In Figure \ref{fig:RSNAPD_results} we compare the offset in deconvolved SFR and velocity dispersions. The left panel of Figure \ref{fig:RSNAPD_results} shows that likely interacting galaxies had positive SFR offsets compared to their controls, $0.14^{+0.10}_{-0.12}$ for the $\lambda_\mu=4$ fits.
The right panel of Figure \ref{fig:RSNAPD_results} compares the $\Delta \log$ offset in deconvolved velocity dispersion from the control. The offsets were calculated using only well-resolved galaxies from the kinematic sample. There was no substantial offset in likely interacting galaxy velocity dispersions from their controls. The choice of velocity regularisation $\lambda_\mu$ did not affect the $\Delta \log$ offsets within the 1-sigma errors for either SFR or velocity dispersion.

\begin{figure*}
    \centering
    \includegraphics[width=\linewidth]{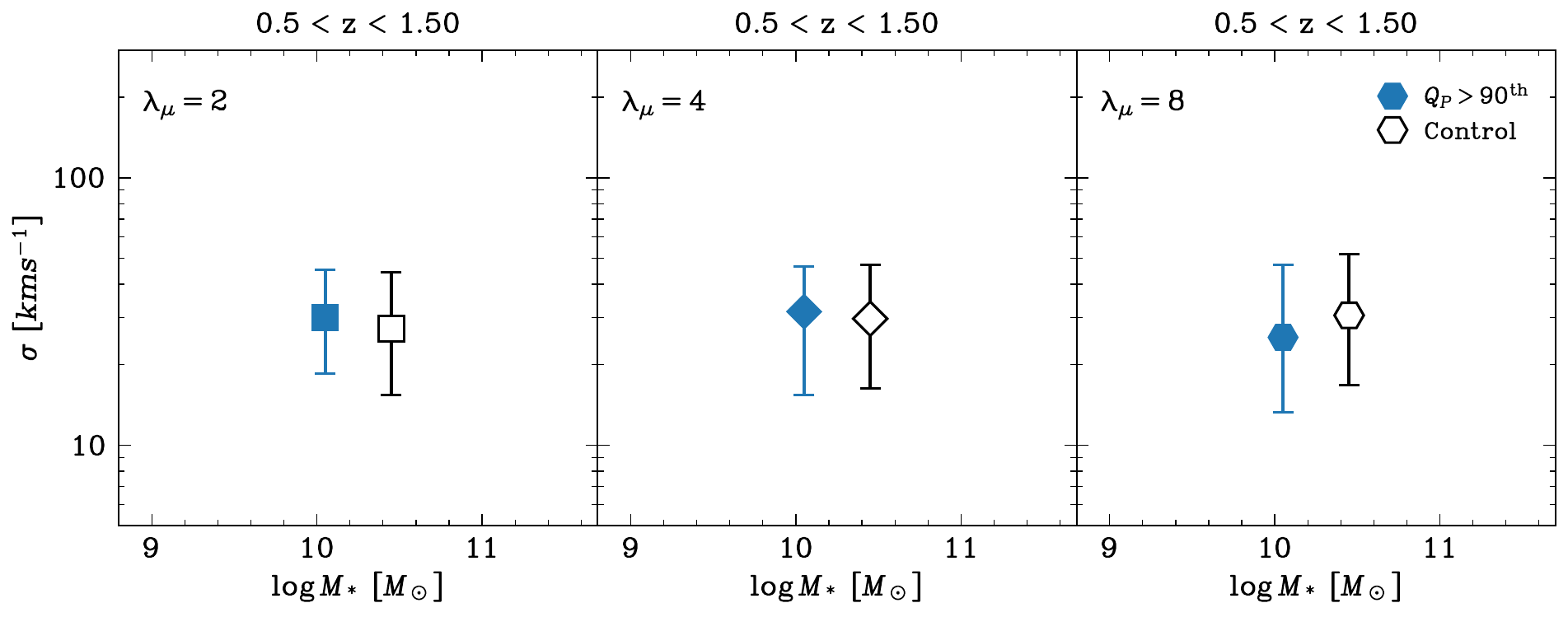}
    \caption{The Kaplan-Meier estimator results of galaxy velocity dispersions for velocity regularisation values of $\lambda_\mu = 2$, $4$ and $8$.
    Only galaxies with redshifts in the range $z=0.5-1.5$ are included.
    The median, $16$\textsuperscript{th} and $84$\textsuperscript{th} percentiles of the Kaplain-Meier estimator are shown by errorbars with closed and open markers for the likely interacting and control samples respectively.}
    \label{fig:RSNAPD_disp_Kaplan_Meier}
\end{figure*}

However, unlike the H$\alpha$ fluxes and SFRs, velocity dispersion measurements included upper bounds. The method used in Figure \ref{fig:RSNAPD_results} treats dispersion upper bounds in the same way as actual measurements, as a result, the upper bounds likely biased the calculated dispersion offsets. In Figure \ref{fig:RSNAPD_disp_Kaplan_Meier} we specifically account for upper bounds using the Kaplan-Meier estimator \citep{Kaplan_1958} with left censoring. The Kaplan-Meier estimator is a non-parametric survival analysis method commonly used in astronomical studies to account for non-detections when calculating dataset statistics \citep{Feigelson_1985}.
We restricted our analysis to galaxies in the range $z=0.5-1.5$, as there was only one likely interacting galaxy with a non-upper bound dispersion in $z-1.5-3.0$.
For the likely interacting and control samples we calculated the median, $16$\textsuperscript{th} and $84$\textsuperscript{th} percentiles of the Kaplain-Meier estimator survival function using the {\tt lifelines} {\tt Python} package \citep{Davidson-Pilon_2019}.
All deconvolved velocity dispersion offsets in the $0.5 < z < 1.5$ bin are within error of their controls, for all regularisation values. There is larger scatter in the $1.5 < z < 3.0$ bin as our sample includes less galaxies in this range, as shown in Figure \ref{fig:nintey_percent_completeness_lim}, though the measured velocity dispersions for the likely interacting and control galaxies remain similar.

We have tested the measured H$\alpha$ flux, $A_{\rm{V}}$, SFR and velocity dispersion offsets using a range of quality cuts (such as increasing the S/N limit or applying stricter ``well-resolved'' size cuts) and varying the size of bounds used to match to control galaxies.
The $\Delta \log$ H$\alpha$ offsets did not change substantially with different quality cuts, and the offsets were consistently positive.
$\Delta \log$ SFR offsets were also consistently positive, though the exact offset we measured was reasonably sensitive to changes in the interacting and control samples, and the median value of the likely interacting sample varied by around $\pm 0.05$ dex depending on the cuts applied. As the $\Delta \log$ H$\alpha$ offsets were relatively stable, the attenuation $A_{\rm{V}}$ values were the likely source of the SFR offset variation.
The velocity dispersion offsets measured using either the $\Delta \log \sigma$ or Kaplan-Meier method did not vary substantially with different quality cuts and were consistent with no offset in velocity dispersion between the likely interacting and control samples.

\section{Discussion}
\label{sec:discussion}

We find an increase in the SFR of likely interacting pairs at $0.6 < z < 2.6$ in the KMOS\textsuperscript{3D} sample, which have projected separations $\lesssim 100$ kpc (depending on galaxy size; see Figure \ref{fig:Q_P_vs_proj_sep}). In contrast, we find no statistical increase in the velocity dispersions of interacting pairs at $0.6 < z < 2.6$ in the KMOS\textsuperscript{3D} sample. We discuss these results in context below.

\subsection{SFR in close interactions}
In Figure \ref{fig:RSNAPD_results} we show a SFR enhancement in our likely interacting sample of $0.14^{+0.10}_{-0.12}$ for the $\lambda_\mu = 4$ fits. As discussed in Section \ref{sec:results}, we consistently observe a positive $\Delta\log$ SFR offset, though the exact offset we measure does vary depending on the quality cuts applied. The likely interacting galaxies have positive attenuation offsets compared to their isolated controls, $0.10^{+0.14}_{-0.06}$ for the $\lambda_\mu = 4$ sample, which is consistent with previous observations and simulations which find that interacting galaxies have higher levels of dust obscuration \citep[e.g.,][]{Yuan_2012}.

\begin{figure}
    \centering
   \includegraphics[width=\linewidth]{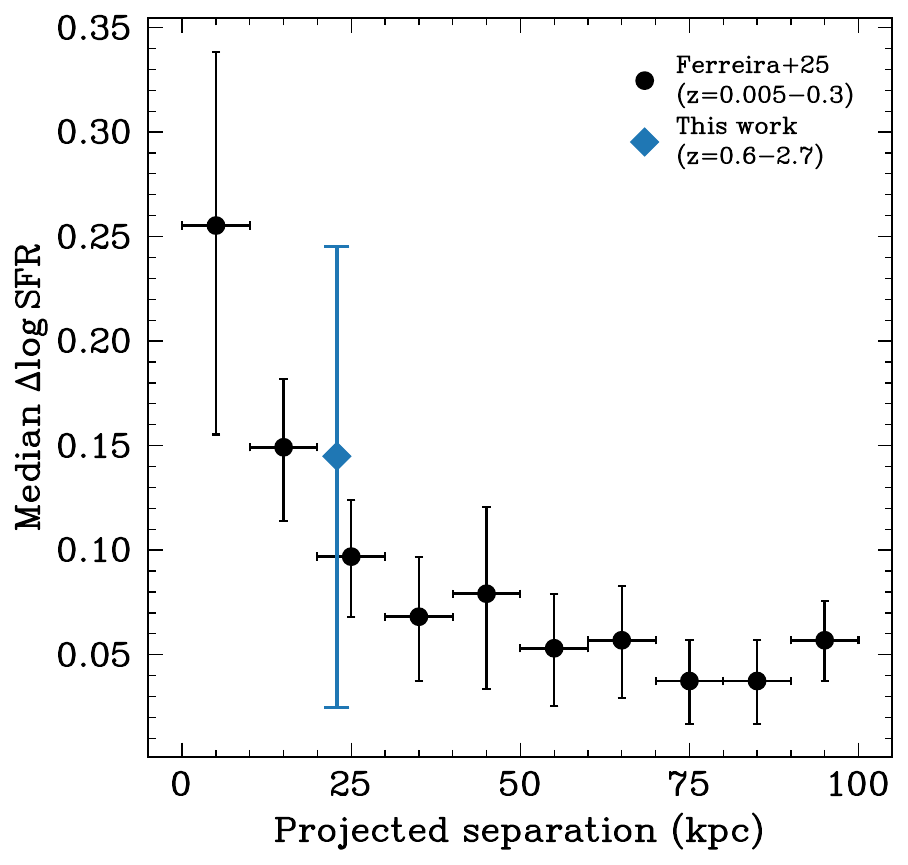}
    \caption{A comparison of the SFR offset of our $\lambda_\mu = 4$ sample at the median $r_{Q_P}$ projected separation of the sample, with offsets from the \citet{Ferreira_2025} pre-merger pair sample ($z=0.005-0.3$) at a range of projected separations.}
    \label{fig:flux_comparison_other_works}
\end{figure}

We find no evolution in SFR offset between our sample at $z=0.6-2.7$ and \citet{Ferreira_2025} at $z=0.005-0.3$. In Figure \ref{fig:flux_comparison_other_works} we include SFR offsets from \citet{Ferreira_2025} of pre-merger close pair galaxies at a range of projected separations (circles). We show the SFR offset of our $\lambda_\mu=4$ sample at $22.9$ kpc, the median $r_{Q_P}$ projected separation of the sample (diamond), which align closely with the offset found by \citet{Ferreira_2025}. The \citet{Ferreira_2025} sample is comprised of $\log(M_\star/M_\odot) > 10$ galaxies with SFRs calculated from aperture-corrected H$\alpha$ measurements. These galaxies form a good comparison to our sample with galaxy masses $\log(M_\star/M_\odot) = 9.2-11.3$ (median $10.19$) and H$\alpha$-based SFR offsets which include star formation from the majority of the optical extent of each galaxy, measured within a $\sim 4\textnormal{"} \times 4\textnormal{"}$ field of view ($\sim 32 \, \textnormal{kpc} \times 32 \, \textnormal{kpc}$ at $z=1.02$, the median redshift of our sample).
Our pair selection method differs from \citet{Ferreira_2025}, who use projected separation and relative velocity cuts of $< 100$ kpc and $\Delta v < 300$ kms$^{-1}$ respectively, though the $r_{Q_P}$ separation of our interacting sample provides a similar comparison. \citet{Ferreira_2025} also select galaxies with pairs with mass ratios $>1$:$10$. Whilst we do not apply a specific pair mass ratio cut in our sample, galaxies with the highest interaction strengths are generally galaxies with more massive companions, as discussed in Appendix \ref{Appendix:merger_mass_ratio}.

There have been a number of studies of SFR enhancement in interacting galaxies at $z > 1$, including new observations with {\tt JWST}  out to $z=8.5$ \citep[e.g.,][]{Silva_2018,Wilson_2019,Horstman_2021,Shah_2022,Calabro_2026,Duan_2026}.
Most works find interacting galaxies have higher SFRs than isolated galaxies, although there is substantial scatter, likely due to the range of approaches (e.g. pair criteria, mass range, SFR measurement technique), which make it difficult to directly compare to our results. Some works also find no increase in SFR enhancement \citep[e.g.,][]{Silva_2018,Wilson_2019}.
The most substantial difference between the studies is the SFR-measurement technique, which can generally be separated into H$\alpha$- and SED-based methods. Most SED-based SFR offset measurements are lower than H$\alpha$-based offsets at the same redshift.

Using UV+IR SFRs from \citep{Whitaker_2014} we can calculate the SED-based SFR offset for our likely interacting galaxies, $0.05^{+0.06}_{-0.04}$ for the $\lambda_\mu = 4$ sample. This is a reduction of around $0.1$ dex from our H$\alpha$-based SFR offsets.
\citet{Horstman_2021} find similar differences between their H$\alpha$- and SED-based SFR offsets and suggest that the difference may be due to the star formation timescales on which the SFR indicators are sensitive. H$\alpha$ SFR measurements are sensitive to star formation on short timescales ($\sim 10$ Myr) whilst the SED-based SFR measurements from \citet{Horstman_2021} are sensitive to star formation on $\sim 100$ Myr timescales. This could suggest that merger-driven SFR enhancements at cosmic noon occur on timescales shorter than $\sim 100$ Myr, which SED-based SFR measurements may average over. SED fitting codes which use non-parametric star-formation-histories may be able to more successfully capture merger-driven SF increases on $\sim 10$ Myr timescales \citep{Calabro_2026}.
However, we note that \citet{Puskas_2025a} only found significant enhancement in sSFR on $\sim 50-100$ Myr timescales in major mergers at $z = 3-9$, and not on $5-10$ Myr timescales.

Interaction-driven SFR offsets have also been explored in simulations. Most studies find increased SF in interacting galaxies, though the observed offsets are sensitive to parameters noted above (e.g. mass ratio), as well as galaxy properties that evolve with time (e.g. gas fraction).
Hydrodynamical simulations of massive discs in $1$:$1$ major mergers generally find significant decreases in SFR offsets as gas fractions increases \citep{Fensch_2017,Scudder_2015}. Using {\tt TNG100} and {\tt TNG300} cosmological simulations which include interactions at a range of mass ratios and gas fractions, \citet{Patton_2020} also find decreases in sSFR offset between $z=0$ and $z = 1$. Yet, other studies utilising the {\tt TNG} and {\tt SIMBA} cosmological simulations to measure SFR offsets in post-merger galaxies do not find significant variations over $0 < z < 1$ \citep{Rodriguez-Montero_2019,Hani_2020}.
Our results show similarities with some simulation SFR offset trends across cosmic time, particularly findings from cosmological simulations which include a range of mass ratios and gas fractions, as these merger samples are more similar to our observed sample, as explored in Appendix \ref{Appendix:merger_mass_ratio}.

\subsection{Velocity dispersion in close interactions}
\label{subsec:discussion_vel_disp}

In contrast to SFRs, we do not measure an enhancement in ionised gas velocity dispersions of close galaxy pairs. In some respects this is surprising.
Observational and simulation studies at $z < 1$ suggest that interactions and merger events can increase the level of turbulence in a galaxy by around $0.1-0.3$ dex, through inducing shocks, gas inflows and increasing star formation rates \citep[e.g.,][]{Rich_2015,Puech_2019,Baron_2024,Fensch_2017,Jimenez-Henriquez_2024}. However, the lack of increased velocity dispersion in the likely interacting galaxies compared to their isolated controls in Figures \ref{fig:RSNAPD_results} and \ref{fig:RSNAPD_disp_Kaplan_Meier} suggests that, despite increasing SFRs, interactions do not significantly increase the level of galaxy turbulence in our sample.

Theoretically, models predict that SFR and turbulent motions of the gas in galaxies should be correlated for discs in quasi-equilibrium \citep[e.g.][]{Krumholz_2016,Krumholz_2018,Ginzburg_2022}. In the models, different mechanisms contribute to maintaining high levels of turbulence including star formation feedback, gas transport, and accretion. Mergers are not included but could induce any of these processes. In the `transport+feedback' model of \cite{Krumholz_2018} and the updated models in \citet{Ginzburg_2022}, a $\Delta\log$SFR increase of $0.14$ would be equivalent to a $\Delta\log\sigma$ increase of $\sim0.13$ for the `high-z' galaxy model. We discuss below the level of dispersion increase that would likely be detectable within our sample. 

Simulations of galaxy mergers find that mergers cause proportionally lower increases in velocity dispersion in gas-rich systems than gas-poor. Idealised hydrodynamical simulations of major mergers find velocity dispersion enhancements of $\Delta \log \sigma \sim 0.60$ in gas-poor mergers but only $\Delta \log \sigma \sim 0.23$ in gas-rich mergers, finding proportionally fewer interaction-driven gas inflows and shocks  \citep{Fensch_2017}.
Similarly, in the larger scale cosmological EAGLE simulations maximum velocity dispersion offsets of $\Delta \log \sigma \sim 0.16$ and $0.22$ are measured in galaxy mergers at $0.1 < z < 0.5$ for low- and high-mass halos respectively (stellar masses of around $\log_{10} M_{*}/M_\odot \in [10.5-11.0]$ and $\log_{10} M_{*}/M_\odot \in [11.0-11.5]$). Velocity dispersion offsets in mergers at $0.7 < z < 1.3$ were less significant, peaking at a median offset of $\Delta \log \sigma \sim 0.12$ for both halo masses \citep{Jimenez-Henriquez_2024}. In both of these studies, the velocity dispersion enhancement was measured at its peak which occurs at either first pericentre passage, second pericentre passage, or coalescence. The peak is typically short lived, $50-100$ Myr or roughly when galaxies are separated by $\lesssim25$ kpc. In between these key timesteps, when the companion is farther away, the enhancement is markedly less. In our work we are averaging over the merger stages with $r_{Q_P}\lesssim100$ kpc and specifically removing mergers near coalescence. Taking these caveats into consideration, our results are not in tension with the simulations but warrant a larger sample with higher spectroscopic completeness.

It is also possible that there is an intrinsic increase in the turbulence of interacting galaxies across the full sample (not just at the smallest separations as discussed above), but we are unable to detect it. The combination of factors such as data quality, analysis method or sample size may mean the uncertainty on our observed dispersion measurements are larger than the intrinsic increase.
The spectral resolution of our sample in particular, limits our ability to detect merger-driven $\Delta \log \sigma$ offsets.
When intrinsic galaxy dispersions are near or below the level of the instrument spectral resolution, deconvolved {\tt ROHSA-SNAP} velocity dispersion measurements have larger errors and are more likely to result in upper bound dispersion measurements (median of $\sigma_{\rm{inst}} = 34.1^{+4.3}_{-2.8}$ kms$^{-1}$, varying between $\sigma_{\rm{inst}} \sim 26$ kms$^{-1}$ and $\sigma_{\rm{inst}} \sim 52$ kms$^{-1}$ in our sample). These values are comparable to dispersions measured in pre- and post-coalescence galaxies in the simulations, particularly for the lower mass halos (e.g. control galaxies had median velocity dispersions of $\sim34$ kms$^{-1}$ for the $\log_{10} M_{200}/M_\odot \in [11.5-12.0]$ halos, which increased to maximum velocity dispersions of $\sim45$ kms$^{-1}$ for the low-halo mass samples in EAGLE). These fractional changes would be difficult to robustly detect in the lower resolution data. In comparison, we would expect to recover increases in some of the higher mass simulated mergers (e.g. from $\sim35$ kms$^{-1}$ to $\sim60$ kms$^{-1}$ during coalescence \citep{Fensch_2017}).

\begin{figure}
    \centering
    \includegraphics[width=\linewidth,trim={0cm 0cm 0cm 3.3cm},clip]{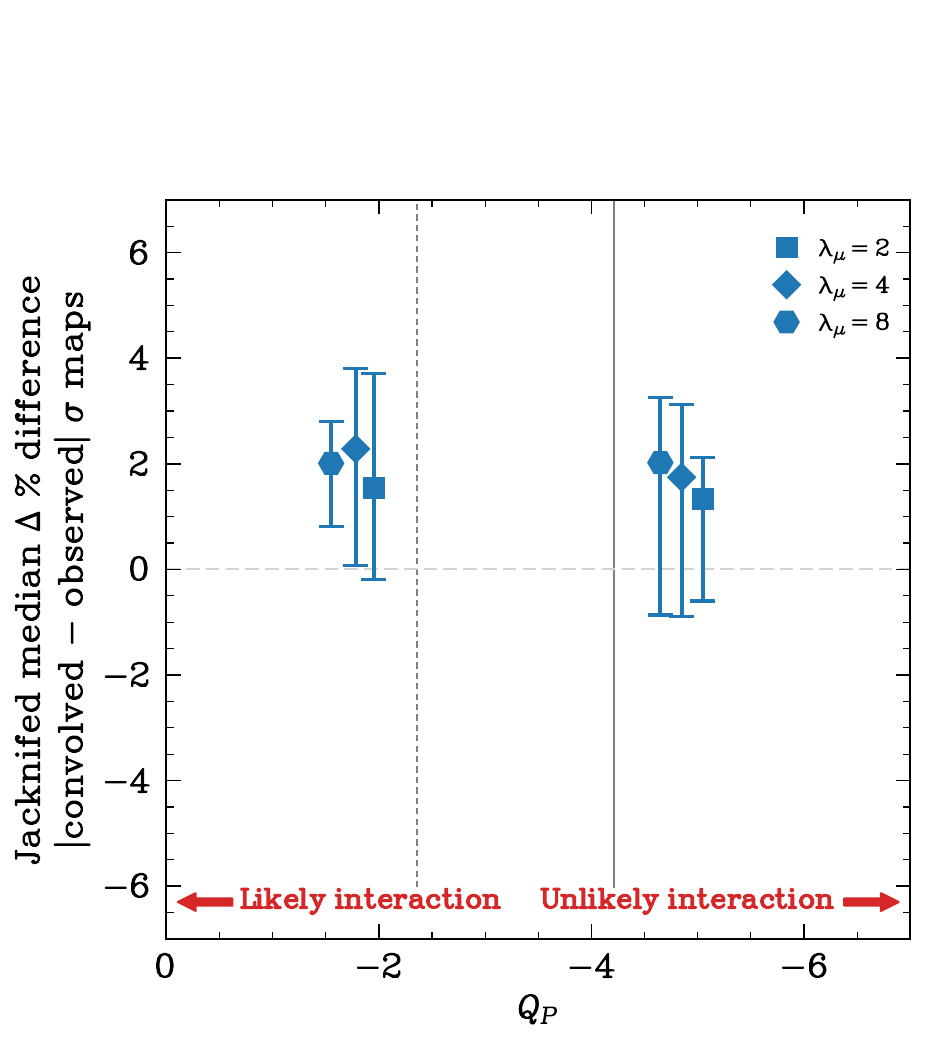}
    \caption{The $\Delta$ offset of the percent difference between the convolved {\tt ROHSA-SNAP} and observed velocity dispersion maps for the kinematic sample. The layout of the figure is similar to Figure \ref{fig:RSNAPD_results_FHa}.}
    \label{fig:RSNAPD_disp_percent_diff}
\end{figure}

The spatial resolution of our data, $\sim4-5$ kpc, is also limiting. We assume spatially-uniform deconvolved velocity dispersions, meaning we average over small-scale variations in the intrinsic dispersions. Spatially varying velocity dispersions have been identified in merging galaxies in the local universe \citep[e.g.,][]{Rich_2015,Pilyugin_2021} and are recoverable with {\tt ROHSA-SNAP} \citep{Kanowski_2025}. If the spatial and spectral resolution of our data were sufficient, we would expect to be able to resolve such fluctuations in our interacting sample. We test if our assumption of a uniform velocity dispersion results in poorer fits to the data in the interacting sample compared to the controls. In Figure \ref{fig:RSNAPD_disp_percent_diff} we show the median percent absolute difference, |model$-$observed|$/$observed$\times100$, of the convolved {\tt ROHSA-SNAP} model and the observed velocity dispersion maps for all spaxels with convolved S/N$>1$, using maps obtained in the same way as Section \ref{sec:results}. The median percent difference of the interacting sample is within error of the control values for all but the $\lambda_\mu=8$ sample, which only has a small percent difference.

Despite these limitations, our results can most likely rule out strong merger-induced velocity dispersion, such as a factor of $2$ ($0.3$ dex) difference, because this would increase velocity dispersions in the likely interacting sample significantly above the spectral resolution, which we would expect to be able to accurately capture using {\tt ROHSA-SNAPD}.
We therefore find similar or lower velocity dispersion offsets in our sample compared to observations at $z\sim 0.6$ of $\Delta\log\sigma \sim 0.18-0.3$ \citep{Puech_2019}, which is in general agreement with results from simulations, that find lower velocity dispersion offsets at higher redshift \citep{Fensch_2017,Jimenez-Henriquez_2024}.

Constraints on velocity dispersion offsets in our sample of $\lesssim 0.3$ also suggest that it is unlikely that mergers alone could drive the increased velocity dispersion of our sample compared to $z\sim0$, or contribute significantly to the scatter in our sample's velocity dispersions at fixed mass and redshift.

High spectral and spatial resolution observations will more closely constrain the level of merger-driven velocity dispersion enhancement at cosmic noon.
The AO-assisted IFS instrument ERIS on the VLT \citep{Davies_2023}, with a spectral resolution of around $\rm{R} = 10 \, 000$ ($\sigma_{\rm{inst}} \sim 13$ kms$^{-1}$), could be used to more accurately measure the velocity dispersion of mergers at cosmic noon, as most galaxies at this epoch have dispersions between $10$ and $100$ kms$^{-1}$. However, larger samples are needed.
Statistical samples of interacting galaxies with high spectroscopic redshift completeness would put finer constraints on both velocity dispersion and SFR enhancements. It would also be possible to compare trends both as a function of interaction strength and for specific subsamples of the population, such as by galaxy mass, redshift or merger mass ratio.
Comparing H$\alpha$ and various timescales of SED-based SFR measurements of the same galaxies, similar to \citet{Horstman_2021,Puskas_2025a,Calabro_2026}, would also provide key information on merger-driven star formation across cosmic time.
Upcoming AO-assisted, multi-plexed IFS instruments such as GIRMOS \citep{Sivanandam_2024} on the Gemini telescope and MOSAIC \citep{Evans_2015} on the Extremely Large Telescope (ELT) and multi-object spectrographs such as MOONS \citep{Cirasuolo_2020} on the VLT will be well suited to provide large samples required for detail future studies. However, as discussed in Section \ref{sec:sample_and_methodology}, care must be taken to avoid contamination from neighbouring galaxies at very close separations, which is particularly important for observations with fibre-fed instruments where spatial masking is not possible.

\section{Conclusions}
We have used the interaction strength parameter $Q_P$ to identify likely interacting galaxies within the KMOS\textsuperscript{3D} survey at cosmic noon. We obtained deconvolved kinematics and fluxes of $131$ star-forming galaxies ($120$ well resolved) using a spatially non-parametric approach and compare the properties of likely interacting galaxies ($Q_P>-2.4$; $10\lesssim r_{Q_P}$kpc] $\lesssim 100$) with isolated controls ($Q_P<-4.2)$ matched in mass and redshift. We find that the likely interacting galaxies have higher H$\alpha$ fluxes ($\Delta\log f_{\mathrm{H}_\alpha}= 0.06^{+0.01}_{-0.03}$) and SFRs ($\Delta\log\mathrm{SFR}=0.14^{+0.10}_{-0.12}$) and are dustier ($\Delta A_{\rm{V}}$=$0.10^{+0.14}_{-0.06}$) than isolated galaxies. Comparing to studies across a range of redshifts, we find no evolution in the SFR offset in $z\sim0$ galaxies and our sample at $z\sim1$.
In contrast to SFR, we find no measurable offset in the velocity dispersions of the interacting sample. This suggests that interactions do not substantially increase turbulence in our sample when averaged over $10\lesssim r_{Q_P}$[kpc] $\lesssim 100$, at least less than a factor of $\sim2$. We note that mergers at their closest interaction are not included in this analysis ($r_{Q_P}$[kpc] $\lesssim10$). Our results are generally consistent with merger simulations and theoretical arguments which suggest that high-redshift, gas-rich mergers cause proportionally lower increases in velocity dispersion than gas-poor mergers. Future observations with high spectral resolution instruments targeting large numbers of interacting galaxies are required to test the robustness of our results. These observations will provide more accurate velocity dispersions measurements and increase population statistics, facilitating the use of finer interaction strength bins and comparisons between specific merger types, such as major vs. minor and gas-rich vs. gas-poor.

\section*{Acknowledgements}
We thank L. Ferreira and S. Ellison for providing their data, which strengthened this work.

\textit{Conflict of interest and funding:}
The authors declare no conflict of interest.
IK acknowledges that this research was supported by an Australian Government Research Training Program (RTP) Scholarship.
N.M.F.S. acknowledges funding by the European Union
(ERC, GALPHYS, 101055023).
Views and opinions expressed are, however, those of the author(s)
only and do not necessarily reflect those of the European Union or
the European Research Council. Neither the European Union nor the
granting authority can be held responsible for them.
TT is supported by the JSPS Grant-in-Aid for Research Activity Start-up (25K23392) and the JSPS Core-to-Core Program (JPJSCCA20210003)

\textit{Software used in this work:} {\tt ROHSA-SNAPD} \citep{Kanowski_2025}, {\tt SMPLOTLIB} \citep{Li_2023_smplotlib_package}, {\tt JAX} \citep{Frostig_2018,jax2018github}, {\tt Scipy} \citep{Virtanen_2020_SciPy}, {\tt Numpy} \citep{Harris_2020}, {\tt Astropy} \citep{Astropy-Collaboration_2013,Astropy-Collaboration_2018,Astropy-Collaboration_2022}, {\tt Matplotlib} \citep{Hunter_2007}, {\tt lifelines: survival analysis in Python} \citep{Davidson-Pilon_2019}, { \tt Uncertainties: a Python package for calculations with uncertainties}, Eric O. LEBIGOT.

\section*{Data Availability}
This work made use of publicly released data and catalogues from KMOS\textsuperscript{3D} \citep{Wisnioski_2015,Wisnioski_2019}, 3D-HST Treasury Survey \citep{Brammer_2012,Skelton_2014,Momcheva_2016}, CANDELS \citep{Kodra_2023}, UVCANDELS \citep{Mehta_2024} and \citet{van_der_Wel_2014}.

Catalogues and kinematic fits created in this work are available on request.



\bibliographystyle{mnras}
\bibliography{bibliography} 




\appendix

\section{Potential biases in the $Q_P$ parameter}
\label{Appendix:Q_P_statistic}

\subsection{Choice of redshift weighting}
\label{Appendix:choice_of_redshift_weighting}
In Equation \ref{Equation:Q_P_full_eqn} we used a redshift weighting term, $w_{iz}$, within an implied velocity range of $\pm 500$ kms$^{-1}$ to calculate interaction strengths of physically nearby galaxies.

The size of the implied velocity range does impact our results, because the range must be a balance between selecting physically close pairs, and integrating redshift PDFs over a substantial range to ensure companions with grism- and photometric-redshifts and weighted sufficiently.
We discuss the impact of changing the implied velocity range to $\pm300$ and $\pm700$ kms$^{-1}$.

Expanding the implied velocity range to $\pm700$ kms$^{-1}$ increased the calculated $Q_P$ value of most galaxies, as the larger velocity range included galaxies at more distant physical separations in the $Q_P$ summation. Expanding the velocity range only had a small impact on the measured $Q_P$ values and derived delta log H$\alpha$ flux and velocity dispersion offsets. There were more significant changes to the observed SFR offset, which was around $0.14$ dex lower than the $500$ kms$^{-1}$ velocity range result. This could be due to the combination of both adding additional wide pairs to the likely interacting sample, and that the SFR offset measurement is sensitive to small changes in the samples, as discussed in Section \ref{sec:results}.

Narrowing the implied velocity range to $\pm300$ kms$^{-1}$ lowered calculated $Q_P$ values, though the magnitude of the change was more substantial than the $500-700$ kms$^{-1}$ velocity range increase. There was a substantial shift of intermediate $Q_P$ galaxies to the control sample. These changes to the likely interacting and control samples had a significant impact on the derived offsets, and we did not measure offsets in either H$\alpha$ flux or SFR in the likely interacting sample. This lack of observed offset may be due to the fact that we have a substantial fraction of galaxies with grism- and photometric-redshifts. When using a small velocity range, the influence on $Q_P$ from neighbours without spectroscopic redshift is substantially reduced, meaning we likely dilute our control sample with interacting galaxies, thereby reducing observed offsets.

\subsection{Merger mass ratios}
\label{Appendix:merger_mass_ratio}
\begin{figure}
    \centering
    \includegraphics[width=\linewidth]{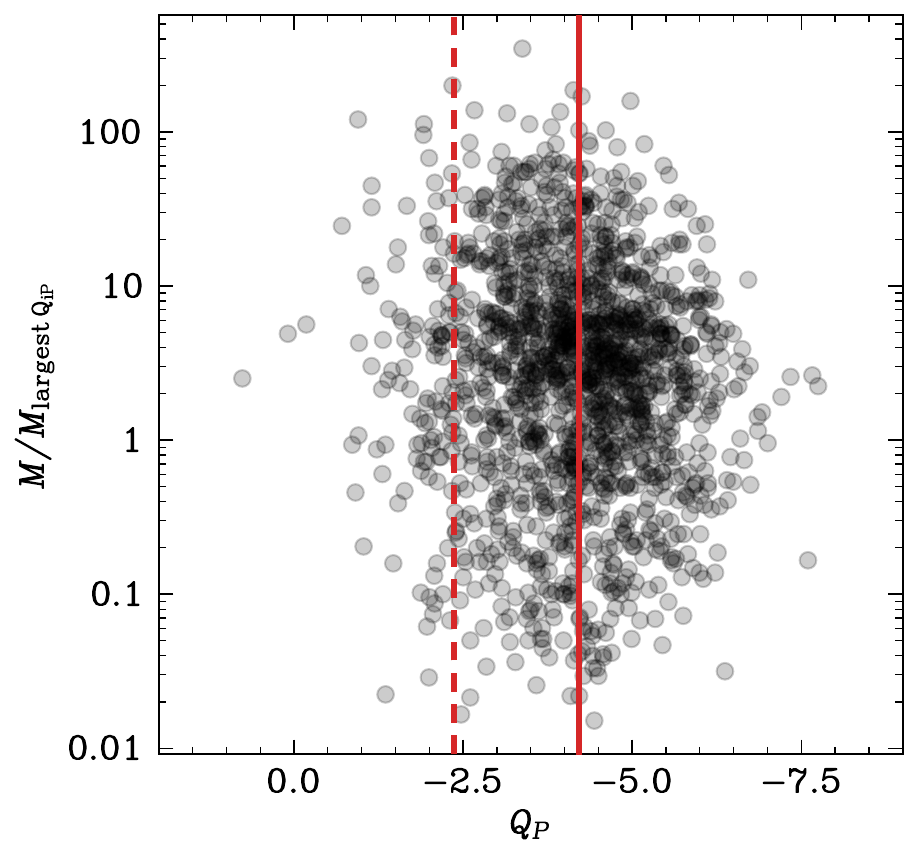}
    \caption{The merger mass ratios of the high-quality 3D-HST spectroscopic sample.}
    \label{fig:3dhst_mass_ratios}
\end{figure}

The $Q_P$ value of a galaxy is primarily influenced by two factors: the relative mass of the galaxy compared to its neighbours', and the galaxy's distance to its neighbours.
We chose to define our interacting sample using the $90$\textsuperscript{th} percentile of the parent sample $Q_P$ because these galaxies are predicted to be most strongly influenced by their neighbours, generally having similar-mass companions at close separations, as demonstrated in Figure \ref{fig:Q_P_vs_proj_sep}, and Figure \ref{fig:3dhst_mass_ratios} which shows the mass ratio of every galaxy in the high-quality 3D-HST spectroscopic sample introduced in Section \ref{sec:results}. The mass ratio of each galaxy was calculated by comparing to the mass of the galaxy which contributed the largest individual interaction strength, $Q_{iP}$, to the galaxy.

\begin{figure}
    \centering
    \includegraphics[width=\linewidth]{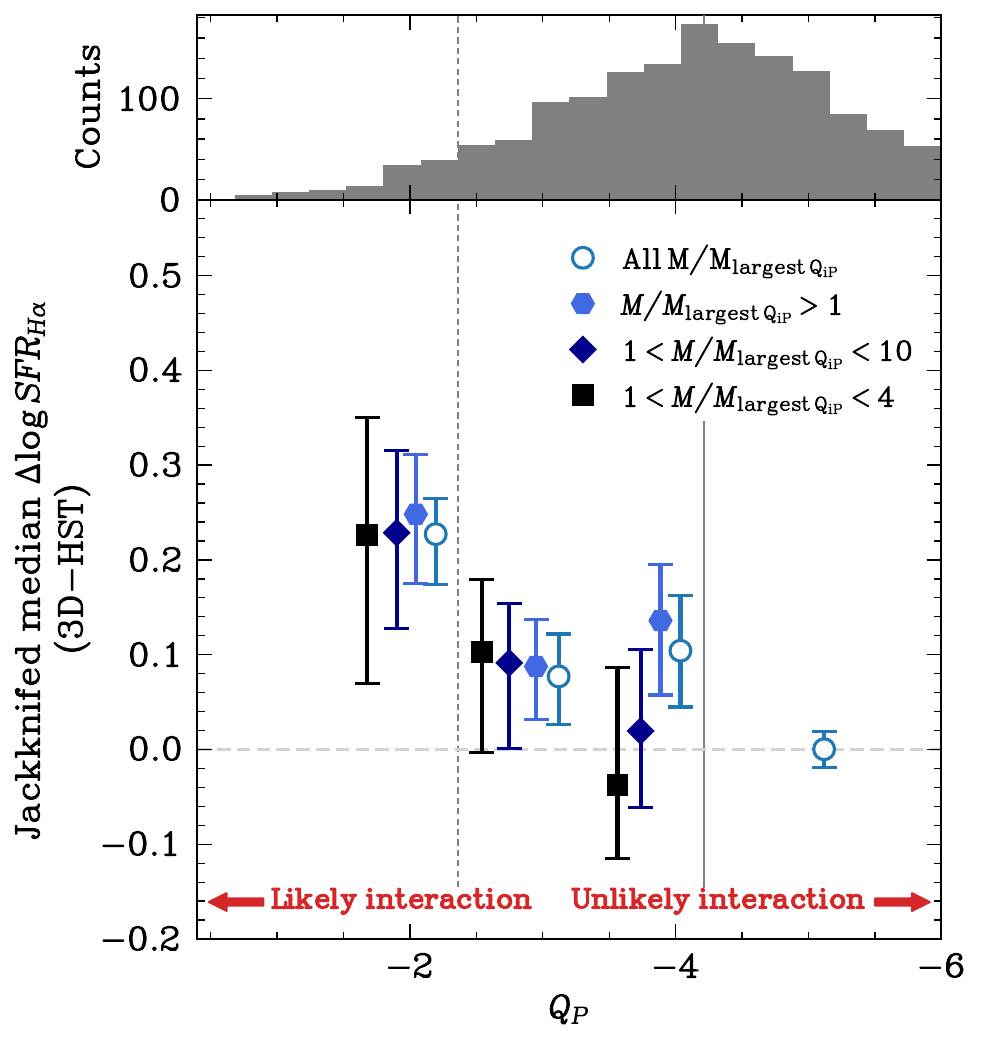}
    \caption{The $\Delta\log$SFR offset of the high-quality 3D-HST spectroscopic sample.
    The offset calculated from all interacting galaxies is shown as an open circle. Offsets calculated using galaxies with specific mass ratios $M/M_{\rm{largest}\,Q_{iP}}$ are shown as closed markers, with shapes denoting different mass ratios.
    The layout of the figure is similar to Figure \ref{fig:RSNAPD_results_FHa}. We also include a histogram showing the distribution of $Q_P$ values.}
    \label{fig:3dhst_dela_log_sfr_mass_ratios}
\end{figure}

It is also possible to study galaxies with lower $Q_P$ values and study trends in merger impacts with interaction strength, in a similar way to previous works which have studied trends in projected separation \citep[e.g.,][]{Ellison_2013}. However, lower $Q_P$ values can be caused by having either less-massive or more-distant neighbours, which complicates observed trends. In Figure \ref{fig:3dhst_dela_log_sfr_mass_ratios} we show the $\Delta \log$ SFR for the high-quality subset of the 3D-HST sample introduced in Section \ref{sec:results}, comparing offsets in galaxies with specific mass ratios. The full sample of galaxies (open points) shows an increasing, but not monotonic, SFR offset with increasing interaction strength. The trend is very similar when only considering galaxies that are the more massive companion $M/M_{\rm{largest}\,Q_{iP}} > 1$ (closed hexagons), though the SFR offsets are slightly increased. This is in-line with other works which find stronger enhancement in the more massive companion \citep[e.g.,][]{Scudder_2012,Davies_2015}. The diamond and square markers show offsets for galaxies with mass ratios between $1 < M/M_{\rm{largest}\,Q_{iP}} < 10$ and $1 < M/M_{\rm{largest}\,Q_{iP}} < 4$ respectively. As the mass ratio range is decreased, the observed trend in $Q_P$ is more strongly influenced by projected separation to the companions than by differences in mass ratio.

To interpret offsets as a function of $Q_P$, mass ratio cuts must be applied. This motivates our choice of interacting sample to be only $Q_P > 90$\textsuperscript{th} percentile; the relatively small kinematic sample size does not allow for further reduction of the sample from mass ratio cuts.

We note that the $\Delta\log$SFR$_{\mathrm{H}\alpha}$ in the most likely interacting sample is marginally higher in Fig.~\ref{fig:3dhst_dela_log_sfr_mass_ratios} than in Fig~\ref{fig:RSNAPD_results}. In the later, galaxies with AGN have been removed, this is not possible in the 3D-HST sample used in Fig.~\ref{fig:3dhst_dela_log_sfr_mass_ratios} and could account for the larger offset.

\subsection{Galaxy effective radii}
\label{Appendix:galaxy_reffs}
\begin{figure}
    \centering
    \includegraphics[width=\linewidth,trim={0cm 0cm 0cm 3.3cm},clip]{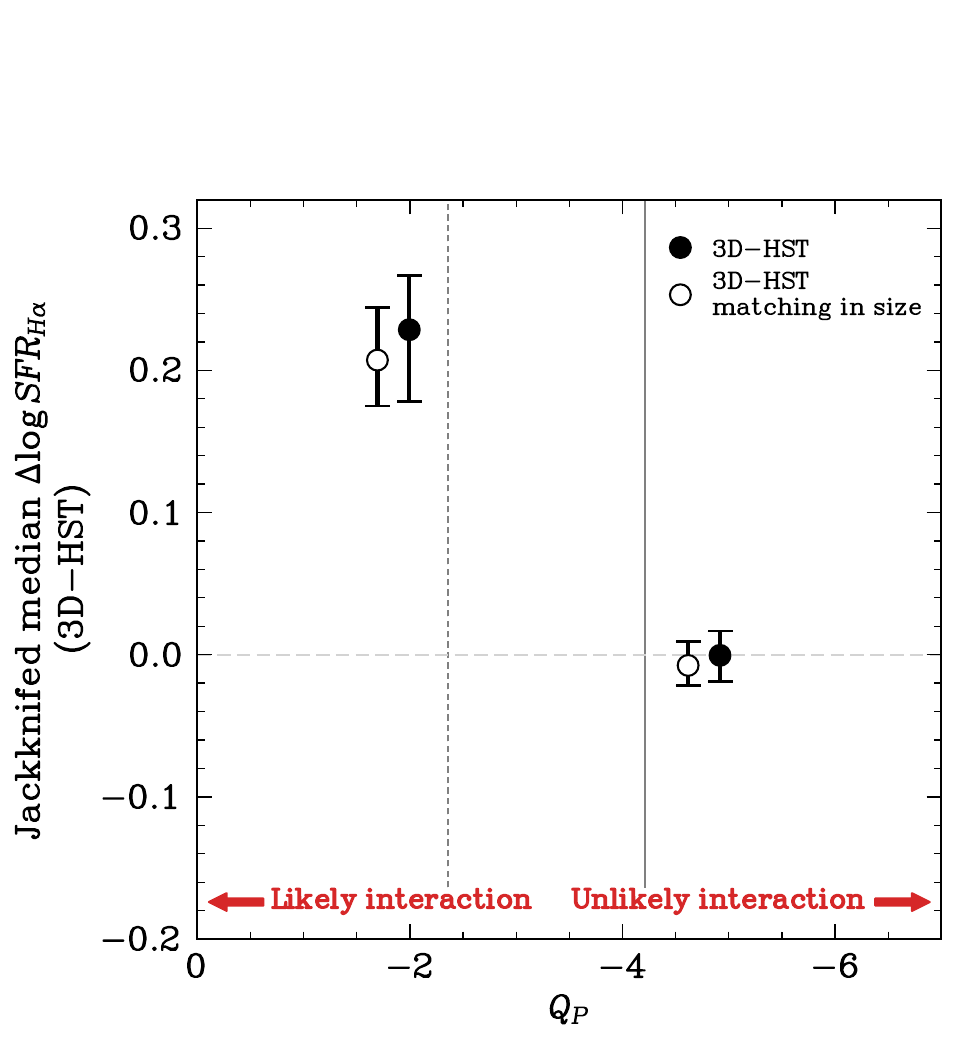}
    \caption{The $\Delta\log$SFR offset of the the high-quality 3D-HST spectroscopic sample. Matching to controls using only mass and lookback time is shown in closed circles, matching also by effective radii is shown in open circles. The layout of the figure is similar to Figure \ref{fig:RSNAPD_results_FHa}.}
    \label{fig:3dhst_matching_reff}
\end{figure}

Figure \ref{fig:Q_P_vs_proj_sep} demonstrates that galaxies with larger effective radii tend to have larger $Q_P$ values. This is because, as described Equation \ref{Equation:Q_iP_derived}, a galaxy with a larger effective radius would experience a greater tidal force at its edge from a given companion distance $D_P$ away than an equal-mass galaxy with a smaller radius.
We checked for the potential impact of differing galaxy sizes by matching galaxy sizes in addition to mass and lookback time.
We matched likely interacting galaxies to controls in a $\pm 0.1$ dex range in effective radii (converted to kiloparsecs), as the size-mass relation in 3D-HST at $0 < z < 3$ has a scatter of $\lesssim 0.2$ dex \citep{van_der_Wel_2014}.
Figure \ref{fig:3dhst_matching_reff} compares the H$\alpha$-based SFR offset of the high-quality 3D-HST spectroscopic sample introduced in Section \ref{sec:results}, when both matching in effective radii (open circles) and when not (closed circles).
There was minimal change in the measured offset of the likely interacting sample when matching by size, $0.207^{+0.037}_{-0.032}$ compared to $0.229^{+0.038}_{-0.050}$ with no matching. The median offset of the size-matched result was slightly lower than the original offset, but still within the $16$th - $84$th percentiles.
From this test, we do not believe that differences in galaxy size have a substantial effect on the results presented in Section \ref{sec:results}. Performing size-matching test on the final kinematic sample itself was not possible as the control sample size was too small to match to a reasonable number of controls.

\section{Updates to {\tt ROHSA-SNAP} }
\subsection{Liikelihood}
\label{Appendix:likelihoods}
We have included the option to use a Cauchy likelihood in the cost function of {\tt ROHSA-SNAPD}\footnote{\url{https://github.com/isaackanowski/ROHSA_SNAPD}}, instead of the default Gaussian likelihood used in \citet{Kanowski_2025}. As \citet{Kanowski_2025}, applied {\tt ROHSA-SNAPD} to mock observations generated from simulated data with added Gaussian noise, a Gaussian likelihood was used through a $\chi^2$ minimisation. However, when applying the code to real observations, a Cauchy likelihood may be more appropriate because it has more power in the wings than a Gaussian, meaning it can more effectively account for outliers, therefore reducing the possibility of the optimiser falling into local minima.

The $L \left(\thetab \right)$ term of the cost function,
\begin{align}
    C \left(\thetab \right) =& \, L \left(\thetab \right) + R \left(\thetab \right).
\end{align}
was originally defined as the least-square difference between the convolved, binned {\tt ROHSA-SNAPD} model, $\tilde{F}_{\textnormal{ binned}}(v_z, \thetab(\rb))$, and the observed data, $F(v_z, \rb)$,
\begin{align}
    L \left(\thetab \right) =& \, \frac{1}{2} \, \sum_{v_z, \, \rb} \left[ \frac{ \tilde{F}_{\textnormal{ binned}}(v_z, \thetab(\rb)) - F(v_z, \rb)}{\bm{\Sigma}(v_z, \rb)} \right]^2 \, ,
\end{align}
where $\bm{\Sigma}(v_z, \rb)$ is the root-mean-square noise for each pixel of $F(v_z, \rb)$. It has now been updated to use the negative log likelihood of a Cauchy distribution,
\begin{align}
    L \left(\thetab \right) =& \, \frac{1}{2} \, \sum_{v_z, \, \rb} \left\{\ln \bm{\Sigma}(v_z, \rb) + \ln \left[ 1 + \left( \frac{ \tilde{F}_{\textnormal{ binned}}(v_z, \thetab(\rb)) - F(v_z, \rb)}{\bm{\Sigma}(v_z, \rb)} \right)^2 \right] \right\} \, .
\end{align}

\section{Kinematics of likely interacting galaxies obtained with {\tt ROHSA-SNAP}}
\label{Appendix:all_Q_P_g_90_fits}
The kinematics maps of all galaxies in the kinematic sample with $Q_P > 90$\textsuperscript{th} percentile will be included in the online version of the published article.


\bsp	
\label{lastpage}
\end{document}